\documentclass[preprintnumbers,superscriptaddress,nofootinbib,aps,prd,floatfix]{revtex4-2}
\usepackage{charter}
\usepackage{footmisc,enumerate}
\usepackage{subfigure}
\usepackage{amsmath,amssymb}
\usepackage{graphicx}
\usepackage{slashed}
\usepackage{xspace,slashed}
\usepackage{hyperref}
\usepackage{setspace}

\usepackage{multirow}
\hypersetup{colorlinks=true, citecolor=blue, urlcolor=blue, linkcolor=blue}
\usepackage[normalem]{ulem}
\usepackage{tikz}
\usepackage{setspace}
\def\be{\begin{equation}}
\def\ee{\end{equation}}
\def\bea{\begin{eqnarray}}
\def\eea{\end{eqnarray}}
\newcommand{\nn}{\nonumber}

\begin{document}
	\setstretch{1.35}
	\title{Frequency space derivation of linear and non-linear memory gravitational wave signals from  eccentric binary orbits}
	
	\begin{abstract}
	The memory effect in gravitational wave (GW) signals is the phenomenon, wherein the relative position of two inertial GW detectors undergoes a permanent displacement owing to the passage of GWs through them. Measurement of the memory signal is an important target for future observations as it establishes a connection between observations with field-theoretic results like the soft-graviton theorems. Theoretically, the memory signal is predicted at the leading order quadrupole formula for sources like binaries in hyperbolic orbits. This can be in the realm of observations by Advanced LIGO, Einstein-Telescope, or LISA for black-holes with masses $\sim$ $O(10^3 \, M_\odot$) scattered by the super-massive black-hole at the galactic center. Apart from the direct memory component there is a non-linear memory signal in the secondary GW emitted from the primary GW chirp-signals emitted by coalescing binaries.  In this paper, we compute the gravitational wave signals and their energy spectrum using the field-theoretic method by computing the scattering amplitudes for eccentric elliptical and hyperbolic binary orbits. The field theoretic calculation gives us the gravitational waveforms of linear and non-linear memory signals directly in the frequency space. The frequency domain templates are useful for extracting signals from the data. We compare our results with other calculations of linear and non-linear memory signals in literature and point out novel features we find in our calculations like the presence of $\log(\omega)$ terms in the linear memory from hyperbolic orbits.
	\end{abstract}
	
	\author{\sf Arpan Hait} \email{arpan20@iitk.ac.in}
	\affiliation{Indian Institute of Technology Kanpur, Kalyanpur, Kanpur 208016, India\\[0.1cm]}
	\author{\sf Subhendra Mohanty} \email{mohantys@iitk.ac.in}
	\affiliation{Indian Institute of Technology Kanpur, Kalyanpur, Kanpur 208016, India\\[0.1cm]}
	\affiliation{Theoretical Physics Division, Physical Research Laboratory, Ahmedabad - 380009, India\\[0.1cm]}
	\author{\sf Suraj Prakash} \email{surajprk@iitk.ac.in}
	\affiliation{Indian Institute of Technology Kanpur, Kalyanpur, Kanpur 208016, India\\[0.1cm]}
	
	\maketitle
	
\section{\bf Introduction}
\noindent The memory effect in gravitational wave (GW) signals is the phenomenon, wherein the relative position of two inertial GW detectors undergoes a permanent displacement owing to the passage of GWs through them \cite{Zeldovich-1, Grishchuk:1989qa, Thorne:1991}. 	The memory effect is categorised broadly according to the nature of the source of GW. Linear memory arises from sources which have unbound components, like hyperbolic binary orbits \cite{Turner-1, Turner-Will}, neutrinos from supernova \cite{Epstein:1978dv, Turner-SN, Kotake:2005zn, Andresen:2016pdt, Morozova:2018glm, Vartanyan:2020nmt, Mukhopadhyay:2021zbt}, gamma-ray bursts \cite{Sago:2004pn,Urrutia:2022lce,Piran:2022gxr} and exotic objects like cosmic strings \cite{Jenkins:2021kcj}.

Another type of memory signal is the non-linear memory \cite{Christodoulou:1991,Will:1991, Blanchet:1992br, Thorne:1992sdb}, where secondary gravitational waves are produced by the primary gravitational waves from sources like coalescing binaries. The importance of non-linear memory is that it will be an experimental proof of graviton-graviton coupling or the non-linear nature of the gravitational  field equations in Einsteinian gravity.

The memory signal has a significance in field theory as it follows from the soft-graviton theorems \cite{Weinberg-0, Weinberg-1, Weinberg-2} where the amplitude of a low energy graviton emission from a scattering process can be related, by a multiplicative kinematic factor, to the hard scattering amplitude without the graviton emission. The zero-graviton frequency amplitude of the soft-amplitude has a pole in the frequency space which in Fourier space is a step function in time which is the characteristic of the memory signal. Weinberg's soft theorem amplitude has been generalised to include higher order terms in graviton momenta using the gravitational gauge invariance and angular momentum conservation \cite{Cachazo:2014fwa}. Calculations of graviton emission amplitudes show that there are non-analytic logarithmic terms in graviton frequencies even in tree level scattering \cite{Laddha:2018myi,Sahoo:2018lxl,Laddha:2018vbn}.

The non-linear memory signal, for binaries in a quasi-circular orbit, already occurs at the 0-PN order \cite{Will:1991, Kenneflick-94}. In \cite{Favata:2008yd, Favata:2010zu} the non-linear memory for the quasi-circular orbit was computed at the 3-PN (Post-Newtonian) order. The non-linear memory signal from eccentric binary orbits was calculated at 3-PN order in \cite{Favata:2011qi} and the 3-PN calculation for eccentric orbits including the tail contributions is done in \cite{Ebersold:2019kdc}. 

Binaries (like the Hulse-Taylor) can have large initial eccentricities but by the time their frequencies enter the threshold of detectors like Advanced LIGO with a threshold of $\sim 10$ Hz, they lose their eccentricities due to gravitational radiation reaction. However, there can be other initial configurations of binary star orbits which could give rise to large eccentricity orbits by the time the frequency enters the Advanced LIGO threshold \cite{Samsing:2017xmd, Tucker:2021mvo}. An eccentric binary can be formed by capture from an unbound orbit \cite{Quinlan, OLeary:2008myb}. An initial three body system may eject one of the bodies and result in an eccentric binary at coalescence \cite{Heggie}. A three body system in which the orbit of one body is at a larger radius, may cause the distant body to perturb the orbit of the tight binary system. This perturbation may give rise to Kozai-Lidov oscillations \cite{Kozai,Lidov} in the binary orbit which can drive the eccentricity to large values \cite{Wen:2002km, Naoz:2012bx}. Constructing gravitational wave templates for coalescing eccentric binaries is therefore of importance for observations \cite{Tucker:2021mvo} at detectors like Advanced LIGO. A determination of eccentricity of the orbits of the coalescing binary will be possible at such detectors by improving the low frequency sensitivity \cite{Favata:2021vhw}. Experimental determination of both the memory signal and the eccentricity of the orbit affect the signal near the low-frequency threshold of the detector, therefore construction of memory signal templates must accurately take into account the eccentricity of the orbit.

The non-linear gravitational wave memory from binary mergers may be discernible once the sensitivity of ground based detectors such as Advanced LIGO \cite{LIGOScientific:2014pky} and Advanced Virgo \cite{VIRGO:2014yos} achieve improved sensitivity in the 5-10 Hz band \cite{Johnson:2018xly, Yang:2018ceq,Yu:2017zgi, Talbot:2018sgr}. The memory effect for individual binaries may be resolvable by forthcoming interferometers such as LISA \cite{LISA:2022kgy}, Cosmic Explorer \cite{LIGOScientific:2016wof} and Einstein Telescope \cite{Einstein-Telescope}. Gravitational-wave memory from binary mergers may also be seen in Advanced LIGO by combining the signals from multiple events \cite{Lasky:2016knh}. The memory signal from several unresolved binary event mergers may be observed as a cumulative change over time in the pulsar timing residuals and may be observed in pulsar timing arrays (PTAs) \cite{ Maiorano:2021sqj, NANOGrav:2019vto, Burke-Spolaor:2018bvk, Rosado:2015epa, Pshirkov:2009ak}.  Ground-based detectors are sensitive to the frequency of the GW in the $5-10^3$ Hz band. Space-based detector LISA  will probe frequencies as low as $10^{-6}$ Hz, and pulsar timing arrays  measure frequencies as low as $10^{-9}$ Hz.

Black-holes in hyperbolic orbits in the gravitational field of super-massive black-holes at the galactic center emanate gravitational wave bremsstrahlung. A $10^3 M_\odot$ black-hole in an hyperbolic orbit around a galactic center black-hole of mass $10^{3} M_\odot$ can produce  gravitational waves signals which can be measured by LISA  \cite{ Capozziello:2008ra, DeVittori:2012da,DeVittori:2014psa,Garcia-Bellido:2017knh, Garcia-Bellido:2017qal,Grobner:2020fnb, Cho:2018upo}. Hyperbolic encounters between massive $(10^2 M_\odot - 10^3 M_\odot)$ black-holes can produces bursts of gravitational waves detectable at Advanced LIGO or Einstein Telescope \cite{Mukherjee:2020hnm, Codazzo:2022aqj}.

For hyperbolic orbits, the nonlinear memory appears at the 2.5-PN  level in the waveform. This is contrary to the case of elliptical and quasi-circular orbits, where the nonlinear memory appears at the Newtonian (0-PN) order \cite{Will:1991, Favata:2011qi}. This is because the radiation reaction effects accumulate over time in the closed orbits while for open orbits the radiation reaction is maximum at the closest approach and is zero at asymptotic past and future times.

The Baysian inference of specific signals from the data is most efficiently achieved by match filtering of signals in the frequency domain with the data \cite{Khan:2015jqa,Garcia-Quiros:2020qpx}. With this aim, we compute the waveforms for the memory signals for eccentric elliptical and hyperbolic orbits in the frequency domain. We compute the waveforms using the tree level graviton emission amplitude with the stress-tensor of the binary orbits as sources. Using this graviton emission amplitude, we compute the frequency spectrum of the energy radiated following \cite{Mohanty:1994yi,KumarPoddar:2019jxe,KumarPoddar:2019ceq,KumarPoddar:2019ceq,Poddar:2021yjd}. For elliptical orbits, the energy radiation spectrum is the source of the non-linear gravitational memory which we thus obtain directly in the frequency space.

The anatomy of this article can be described as follows: In section \ref{sec:QFT-appr}, we outline how the gravitational waveform can be constructed in a field-theoretic approach and we also highlight the computation of the rate of energy loss associated with gravitational wave radiation. In section \ref{sec:hyperbolic}, we compute components of the stress-tensor, in frequency domain, for binaries in hyperbolic orbits. This is followed by the evaluation of the same quantities in the limit of vanishing frequency and the associated linear memory in section \ref{sec:linear-memory}. In section \ref{sec:soft-theorem-comparison}, we highlight how the linear memory waveforms constructed using our approach compare against those obtained based on soft-graviton theorems. Next, we elucidate the steps involved in constructing the non-linear memory waveforms associated with the radiation from binaries in elliptical orbits in section \ref{sec:elliptical}. The general formalism for the non-linear memory effect, and the specific case of circular orbits has been described in the appendices. Calculations for the specific case of binaries in elliptical orbits appear in section \ref{sec:ellip-non-lin-mem}. We summarize our conclusions in section \ref{sec:conclusion}. 

Throughout the paper, we use the natural units with $\hbar=c=1$ and Newtons constant $G=M_{\text{pl}}^{-2}$ where $M_{\text{pl}}$ is the Planck mass with the value $M_{\text{pl}}=1.22 \times 10^{19} {\rm GeV}$.  
\section{\bf Gravitational waveform and energy radiated from scattering amplitudes}\label{sec:QFT-appr}

\noindent The probability amplitude of emitting a graviton of polarisation $\epsilon^\lambda_{\mu \nu}(\vec n)$ from a source with stress-tensor (in the momentum space ) ${\widetilde T}^{\mu \nu} (k)$ is given by
\begin{eqnarray}\label{amp-1}
 {\cal A}_\lambda(k_0, \vec n k_0)
 =-\iota\,\cfrac{\kappa}{2}\,  \epsilon_{\mu \nu} ^{*\lambda}(\vec n)\,
{\widetilde T}^{\mu \nu}(k_0,  \vec n \, k_0 ) \,.
\end{eqnarray}
We can express the gravitational wave metric observed at the detector in terms of the probability amplitude of a graviton emission by a source at a distance $r$ as,
\begin{eqnarray} \label{Greens-3}
h_{\alpha \beta}(\vec x, t)= \cfrac{1}{4 \pi r} \, \int \, \cfrac{d k_0}{(2 \pi)}\, \sum_{\lambda=1}^2\, \epsilon_{\alpha \beta} ^{\lambda}(\vec n)\, {\cal A}_\lambda(k_0, \vec n\, k_0)\, e^{-\iota k_0 (t-r) }\,.
\end{eqnarray}
The graviton field in Eq.~\eqref{Greens-3} is a canonical spin-2 field with mass dimension 1 as it is defined as an expansion of the metric $g_{\mu \nu}=\eta_{\mu \nu} + \kappa\, h_{\mu \nu}$ where $\kappa=\sqrt{32 \pi G}$. The metric perturbation identified as gravitational wave is the dimensionless quantity $\widetilde h_{\mu \nu}\equiv g_{\mu \nu}- \eta_{\mu \nu}= \kappa\, h_{\mu \nu}$. The expression for the dimensionless gravitational wave in terms of the amplitude is therefore, from Eq.~\eqref{Greens-3}, given by 
\begin{eqnarray}\label{Greens-4}
\widetilde h_{\alpha \beta}(\vec x, t)= \cfrac{\kappa}{4 \pi r} \,\int\, \cfrac{d k_0}{(2 \pi)} \, \sum_{\lambda=1}^2 \, \epsilon_{\alpha \beta} ^{\lambda}(\vec n) \, {\cal A}_\lambda(k_0, \vec n\, k_0)\, e^{-\iota k_0 (t-r) } \,.
\end{eqnarray}
This relates the waveform at the detector to the probability amplitude of graviton emission by the source. To relate the waveform at the detector to the source stress-tensor we substitute in Eq.~\eqref{Greens-4}, the expression for the amplitude given in Eq.~\eqref{amp-1}, to obtain
\begin{eqnarray}\label{Greens-5}
\widetilde h_{\alpha \beta}(\vec x, t)&=& -\cfrac{\kappa^2}{8 \pi r} \, \int \,  \cfrac{dk_0}{2\pi} \, \sum_{\lambda=1}^2 \, \epsilon_{\alpha \beta} ^{\lambda}(\vec n) \,  \epsilon_{\mu \nu} ^{*\lambda}(\vec n)\,
{\widetilde T}^{\mu \nu}(k_0,  \vec n\, k_0 )\,  e^{-\iota k_0 (t-r) }\, \nn\\
&=&-\cfrac{4G}{ r} \,\int \,\cfrac{dk_0}{2\pi} \,\left(\widetilde T_{\alpha \beta}(k_0,  \vec n\, k_0 ) - \cfrac{1}{2}\, \eta_{\alpha \beta}\, {\widetilde T^\mu}_\mu (k_0,  \vec n\, k_0 ) \right) \, e^{-\iota k_0 (t-r) }\,.
\end{eqnarray}
where we have made use of the completeness relation 
\begin{eqnarray}\label{Polsum-mu}
\sum_{\lambda=1}^2\, \epsilon_{\mu\nu}^\lambda(k)\,\epsilon_{ \alpha\beta}^{*\lambda}(k) = \cfrac{1}{2}\,(\eta_{\mu\alpha}\,\eta_{\nu\beta}+\eta_{\mu\beta}\,\eta_{\nu\alpha})- \cfrac{1}{2}\,\eta_{\mu\nu}\,\eta_{\alpha\beta}.
\end{eqnarray}
To obtain the propagating degrees of freedom, we need to project the transverse-traceless (TT) components of the wavefunction constructed in Eq.~\eqref{Greens-5}.  
\begin{eqnarray}
\Big[\widetilde h_{ij}\Big]^{\text{TT}}(\vec x, t) = -\cfrac{4G}{ r}\,\Lambda_{ij,kl}(\vec n) \, \int \,  \cfrac{dk_0}{2\pi} \, \left(\widetilde T_{k l}(k_0,  \vec n\, k_0 ) - \cfrac{1}{2}\, \eta_{k l}\, { \widetilde T^\mu}_{\,\,\,\,\mu} (k_0,  \vec n\, k_0 ) \right)\, e^{-\iota k_0 (t-r) }.
\end{eqnarray}
where, $ \Lambda_{ij,kl}(\vec n)$ is the transverse-traceless projection operator defined with respect to the direction, $\hat n$, of the emitted gravitational wave. The explicit form of the TT projection operator is
\begin{eqnarray}\label{tensor-1} 
	\Lambda_{ij,kl}( \hat n)&=&P_{ik}( \hat n) P_{jl} ( \hat n)- \cfrac{1}{2} P_{ij}( \hat n)P_{kl} ( \hat n) = \left(\delta_{ik}-  n_i  n_k\right)\left(\delta_{jl} - n_j  n_l\right)-\cfrac{1}{2} \left(\delta_{ij}- n_i  n_j \right)\left( \delta_{kl}- n_k  n_l\right).
\end{eqnarray} 
Since $\Lambda_{ij,kl} \,\,\eta_{k l}=0$, the ${\widetilde T^\mu}_{\,\,\,\,\mu}$ term vanishes and we obtain the simpler result,
\begin{eqnarray}\label{GWTmunu}
\Big[\widetilde h_{ij}\Big]^{\text{TT}}(\vec x, t)= -\cfrac{4G}{ r}\,\Lambda_{ij,kl}(\vec n)\, \int \,  \cfrac{dk_0}{2\pi}\, T_{k l}(k_0,  \vec n\, k_0 ) \,  e^{-\iota k_0 (t-r) }.
\end{eqnarray}
 In frequency space, the observed gravitational wave and the stress-tensor of the source can be related as,
\begin{eqnarray}\label{GWTmunu-f}
\Big[\widetilde h_{ij}\Big]^{\text{TT}}(\vec x, k_0)= -\cfrac{4G}{ r}\, \Lambda_{ij,kl}(\vec n)\, T_{k l}(k_0,  \vec n\, k_0 ).  
\end{eqnarray}
We shall use Eq.~\eqref{GWTmunu-f} to compute the gravitational waveform from various sources, like compact binaries in bound and unbound orbits, by computing the stress-tensor of the source in frequency space.
\subsection{\bf Power spectrum of gravitational wave in field theoretic approach}
\noindent The rate of graviton emission is  given by the Fermi Golden Rule, and is the amplitude squared summed over the final state graviton polarisation and integrated over the phase space volume,
\begin{eqnarray}\label{dgamma0}
\Gamma &=&\sum_\lambda\, \int\, \cfrac{|S_{fi}|^2}{T} \,  
\cfrac{d^3 \vec k}{(2 \pi)^3\, 2\omega} = \sum_n \,\sum_\lambda \, \int\, \big\vert  {\cal A}_\lambda(\omega, \omega^\prime_n)  \big\vert^2 \,(2 \pi)\, \delta(\omega-\omega_n^\prime)\,
 \cfrac{d^3 \vec k}{(2 \pi)^3\, 2 \omega}\, \nn\\
&=&\cfrac{ \kappa^2}{4}\, \sum_n \, \sum_\lambda \, \int \,  \big\vert T_{\mu \nu}( \vec k, \omega_n^\prime) \, \epsilon_\lambda^{*\mu \nu} (\vec k) \big\vert^2 \,(2 \pi)\, \delta(\omega-\omega_n^\prime)\,
 \cfrac{d^3 \vec k}{(2 \pi)^3\, 2\omega_k}.
\end{eqnarray}
\noindent Here, $\lambda$ accounts for the possible states of polarization and $n$ denotes the harmonics corresponding to the emission. The  energy radiated is obtained from the probability of radiation given above by including a factor of $\omega=|\vec k|$ in the integral,
\begin{eqnarray}\label{dEdt-1}
E_\text{gw}&=&\cfrac{ \kappa^2}{4}\,  \sum_{\omega^\prime_n} \sum_\lambda \, \int \,\big\vert T_{\mu \nu}( \vec k, \omega_n^\prime)\, \epsilon_\lambda^{*\mu \nu} (\vec k) \big\vert^2 \,(2 \pi)\, \delta(\omega-\omega_n^\prime)\,  \,\, \omega \,\,\cfrac{d^3 \vec k}{(2 \pi)^3\, 2 \omega}.
\end{eqnarray}
\noindent The modulus squared piece of the integrand in Eq.~\eqref{dEdt-1} can be simplified using the polarisation sum relation, see Eq.~\eqref{Polsum-mu}, as follows:
\begin{eqnarray}\label{Tmunusquare}
\sum_\lambda \,  \big\vert T_{\mu \nu}( \vec k, \omega_n^\prime)\, \epsilon_\lambda^{*\mu \nu} (\vec k) \big\vert^2 &=& \sum_\lambda \, \left( T_{\mu \nu}( \vec k, \omega_n^\prime)\, T_{\alpha  \beta}^*( \vec k, \omega_n^\prime)\right)\left(\epsilon_\lambda^{*\mu \nu} (\vec k) \, \epsilon_\lambda^{\alpha  \beta} (\vec k) \right)\, \nn\\
&=&T_{\mu \nu}( \vec k, \omega_n^\prime)\, T^{*\nu \mu}( \vec k, \omega_n^\prime) -\cfrac{1}{2}\, \big\vert  {T^\mu}_{\mu}( \vec k, \omega_n^\prime) \big\vert^2\,.
\end{eqnarray}
The $T^{00}$ and $T^{i0}$ components can be expressed in terms of the $T^{ij}$ ones by utilizing
the conserved current relation, $k_\mu T^{\mu\nu}=0$. This allows us to write
\begin{equation}\label{app6}
T_{0j}=-\hat{k^i}\,T_{ij},\hspace{0.5cm} T_{00}=\hat{k^i}\,\hat{k^j}\,T_{ij}.
\end{equation}
Using these relations, we can rewrite Eq.~\eqref{Tmunusquare} as
\begin{eqnarray}\label{app7}
\big\vert T_{\mu\nu}(\vec k, \omega_n^\prime)\big\vert^2 -\cfrac{1}{2}\, \big\vert T^{\mu}{}_{\mu}(\vec k, \omega_n^\prime)\big\vert^2 &=& T_{ij}T^{*ji}+T_{00}T^{*00}+ T_{0i}T^{*i0}+ T_{i0}T^{*0i} - \cfrac{1}{2}\left({T^0}_0+{T^i}_i \right)\left({T^{*0}}_0+{T^{*j}}_j \right)\, \nn\\
&=&\left(T_{ij}T^{*ji} -\cfrac{1}{2}\, T^i_i T^{*j}_j\right) +\cfrac{1}{2} \,\hat{k^i}\hat{k^j}\hat{k^l}\hat{k^m} T_{ij} T^*_{lm} - \left( \hat{k^l}\,\hat{k^m}\, T_{il} T^*_{mi} + \hat{k^l}\,\hat{k^m}\, T_{il} T^*_{mi}\right)
\nn\\
&& \qquad\,+\,  \cfrac{1}{2}\,\left(\hat{k^l}\,\hat{k^m} \,T^*_{lm} T^{i}_i + \hat{k^l}\,\hat{k^m} \, T_{lm} T^{*j}_j \right).
\end{eqnarray}
In the quadrupole approximation of the source, for sources smaller in size than the wavelength of the GWs, $\vec k \cdot \vec x \ll 1$, the stress-tensor in momentum space $T_{\mu\nu}( \vec k, \omega_n^\prime)$ has no explicit $\vec{k}$ dependence. Therefore, after substituting the contents of Eq.~\eqref{app7} in Eq.~\eqref{dEdt-1}, one can perform the angular integrations using the following relations:
\begin{eqnarray}\label{dOmegak}
&&\int d\Omega_k =4 \pi,  \hspace{0.9cm}\int d\Omega_k \,\hat{k^i}\,\hat{k^j}=\cfrac{4\pi}{3}\,\delta_{ij},   \hspace{0.7cm} \int d\Omega_k \, \hat{k^i}\,\hat{k^j}\,\hat{k^l}\,\hat{k^m}=\cfrac{4\pi}{15}\,\left(\delta_{ij}\,\delta_{lm}+\delta_{il}\,\delta_{jm}+\delta_{im}\,\delta_{jl} \right).
\end{eqnarray}
to obtain 
\begin{equation}\label{app9}
\int\, d\Omega_k \, \Big[|T_{\mu\nu}(\vec k, \omega_n^\prime)|^2-\cfrac{1}{2}\,|T^{\mu}{}_{\mu}(\vec k, \omega_n^\prime)|^2\Big]
=\cfrac{8\pi}{5}\,\Big(T_{ij}(\omega_n^\prime)\,T^*_{ji}(\omega_n^\prime)-\cfrac{1}{3}\,|T^{i}{}_{i}(\omega_n^\prime)|^2\Big),
\end{equation}
Finally, using the result of Eq.~\eqref{app9}, the expression for the energy radiated by a source in terms of the source stress-tensor can be obtained as a modification of Eq.~\eqref{dEdt-1} as
\begin{eqnarray}\label{dEdt-2}
E_\text{gw}=\cfrac{ \kappa^2}{4} \, \sum_{\omega^\prime_n}\,  \int \,  \cfrac{8 \pi}{5}\,\Big(T_{ij}(\omega_n^\prime)T^*_{ji}(\omega_n^\prime)-\cfrac{1}{3}\,|T^{i}{}_{i}(\omega_n^\prime)|^2\Big)\,{\omega}^3\, 2\pi \,\delta(\omega_n^\prime-\omega) \cfrac{d\omega}{(2 \pi)^3\, 2 \omega} .
\end{eqnarray}
\noindent We make use of this expression for computing the energy radiated by binaries in elliptical orbits which is the source of the non-linear memory signal. We will also show using this expression that in the memory signal from hyperbolic orbits, there is non-zero energy radiated around the zero-frequency band.
\section{Gravitational radiation from hyperbolic binary encounter}\label{sec:hyperbolic}
\noindent For the case of unbound orbits, we consider a black-hole of mass $m_2$ in a hyperbolic orbit around a larger black-hole of mass $m_1$. In the centre of mass frame, an equivalent description is in terms of the motion of single body having the reduced mass $\mu = \cfrac{m_1\,m_2}{m_1+m_2}$. The following quantities describe the system:
\begin{itemize}
	\item The coordinates parametrized as,
\begin{eqnarray}\label{cord:hyp}
			x(\xi) = a\,(e - \cosh\xi),\qquad y(\xi) = b\,\sinh\xi, \qquad z(\xi) = 0, \qquad \cfrac{\omega^\prime}{\nu}\,t = \omega_0\, t = (e\sinh\xi - \xi),
		\end{eqnarray}
 where, $a$ and $b \,\,(= a \sqrt{e^2 - 1},\,\, e > 1)$ denote the semi-major and semi-minor axes and the variables $e$ and $\xi \in (-\infty, \infty)$ refer to the eccentricity and the hyperbolic anomaly of the orbit respectively.
		\item The angular frequency $\omega^\prime$ is proportional to the fundamental frequency $\omega_0$, as $\omega^\prime = \nu \,\omega_0$, where $\nu \in (0, \infty)$ is a non negative real number and $\omega_0 = \left(\cfrac{G\, (m_1 + m_2)}{a^3}\right)^{1/2}$.
	\end{itemize}
\noindent We start with the computation of the stress-tensor components in frequency space. Focussing first on the $xx$ component, the calculation proceeds as follows:	
\begin{eqnarray}
		T_{xx}(\omega^{\prime})&=& \int_{-\infty}^{\infty} dt \,e^{i \omega^\prime t} \,\mu \,\dot x^2 \,= \, -\iota \, \mu \, \omega^\prime \int_{-\infty}^{\infty} dt \,e^{i \omega^\prime t} \, \,x\, \dot x \,= \, -\iota \,\mu \,\omega^\prime \int_{-\infty}^{\infty} d\xi \,e^{i \nu(e \sinh \xi -\xi)} \, \,x\, \cfrac{ dx}{d\xi},
	\end{eqnarray}
where, the second equality is obtained after implementing integration by parts and neglecting a term proportional to $\ddot{x}(t)$\footnote{While the acceleration becomes non-negligible as the point of closest approach is reached, for the rest of the integration domain, the contribution from the $\ddot{x}(t)$ term to the integral is negligible.}. The final expression is obtained after incorporating a change of variables $t \rightarrow \xi$. Since, the integration is over $\xi$, the result will simply be a function of $\nu$. 
Thus we obtain,
	\begin{eqnarray}\label{eq:Txx-hyperb}
		T_{xx}(\omega^\prime)	&=& \, \iota \,\mu \, \nu\, \omega_0 \, a^2 \int_{-\infty}^{\infty} d\xi \,e^{\iota \nu(e \sinh \xi -\xi)} \, \, \sinh \xi\,\, (e- \cosh\xi) \,  \nonumber\\
		&=& \mu\,\nu\, \omega_0\, a^2 \pi\, \left[ \, \cfrac{\iota}{\nu e^2}\, H^{(1)}_{\iota\nu}(ie\nu) - \left(e - \cfrac{1}{e}\right){H_{\iota\nu}^{(1)}}^\prime(\iota e\nu)\, \right].
	\end{eqnarray}
\noindent We replaced $\sinh \xi$, $\cosh \xi$ by their exponential counterparts. Subsequently, we identified and replaced the integrals with Hankel functions, i.e., 
	\begin{eqnarray}\label{eq:Hankel-1}
		H_p^{(1)}(q)= \cfrac{1}{\iota \pi}\int\limits_{-\infty}^\infty d\xi \,\,e^{q\sinh \xi -p\xi} \,\,.
	\end{eqnarray}
Then, we utilized the following recurrence relations for Hankel functions to simplify the expression:
\begin{eqnarray}\label{eq:Hankel-recurr-rel}
		H_{p-1}^{(1)}(q) + H_{p+1}^{(1)}(q) &=& \cfrac{2\,p}{q}\,H_{p}^{(1)}(q)\,\qquad \text{and} \qquad
		H_{p-1}^{(1)}(q) - H_{p+1}^{(1)}(q) = 2\, {H_{p}^{(1)}}^\prime(q).
\end{eqnarray}
	The remaining non-zero components can similarly be evaluated as:	 
	\begin{eqnarray}\label{eq:Tyy-xy-hyperb}
		T_{yy}(\omega^\prime) &=& \int_{-\infty}^{\infty} dt \,e^{\iota \omega^\prime t} \,\mu \,\dot y^2 \,=\, -\iota \,\mu\,\omega^\prime \int_{-\infty}^{\infty} dt \,e^{\iota \omega^\prime t} \, \,y\, \dot y \,= \, -\iota \,\mu \,\omega^\prime \int_{-\infty}^{\infty} d\xi \,e^{\iota \nu(e \sinh \xi -\xi)} \, \,y\, \cfrac{ dy}{d\xi}\nonumber\\
		&=&  \mu\,\nu\, \omega_0\, a^2 (e^2-1)\, \pi \bigg[ \cfrac{\iota}{\nu e^2}\,{H_{\iota\nu}^{(1)}}(\iota e\nu) + \cfrac{1}{e}\,{H_{\iota\nu}^{(1)}}^\prime(\iota e\nu) \bigg], \nonumber\\
		T_{xy}(\omega^\prime) &=& \int_{-\infty}^{\infty} dt \,e^{\iota \omega^\prime t} \,\mu \,\dot x\,\dot y \,=\, -\iota\, \mu\, \omega^\prime \int_{-\infty}^{\infty} dt \,e^{\iota \omega^\prime t} \, \,y\, \dot x \,=\, -\iota\,\mu\,\omega^\prime \int_{-\infty}^{\infty} d\xi \,e^{\iota \nu(e \sinh \xi -\xi)} \, \,x\, \cfrac{ dx}{d\xi} \nonumber\\	
		&=& -\mu\, \nu\, \omega_0\,a^2\,\sqrt{e^2-1}\, \pi\left[ \left( \cfrac{1}{ e^2} - 1 \right)\, {H_{\iota \nu}^{(1)}}(\iota e\nu)  + \cfrac{\iota}{\nu e}\,{H_{\iota \nu}^{(1)}}^\prime(\iota e\nu)\right].
	\end{eqnarray}
Using the expressions for $T_{xx}$, $T_{yy}$ and $T_{xy}$ derived in Eqs.~\eqref{eq:Txx-hyperb} and \eqref{eq:Tyy-xy-hyperb} we obtain,
	\begin{eqnarray}
	T_{ij}(\omega^\prime)\,T^*_{ji}(\omega^\prime)-\cfrac{1}{3}\,|\,T^{i}{}_{i}(\omega^\prime)\,|^2 	&=& \mu^2\,\nu^2\,\omega_0^2\,a^4\,\pi^2\,\Big[\,f_1(\nu,e)\,\left(H_{\iota \nu}^{(1)}(\iota e\nu)\right)^2 + f_2(\nu,e)\,\left({H_{\iota \nu}^{(1)}}^\prime(\iota e\nu)\right)^2 \,\Big].
	\end{eqnarray}
	Here, 
	\begin{eqnarray}\label{eq:f-nu-e}
		f_1(\nu,e) 
		&=& \cfrac{2}{e^4}\,(e^2-1)^3 + \cfrac{6 - 6e^2 + 2e^4}{3\,\nu^2\,e^4} \qquad \text{and} \qquad f_2(\nu,e) = 2\,\left(\cfrac{e^2-1}{e}\right)\left(\cfrac{1}{\nu^2\,e} + \cfrac{e^2-1}{e}\right). 
	\end{eqnarray}
 In Eq.~\eqref{dEdt-2}, we replace the sum over $\omega_n^\prime$ with an integral over $\omega^\prime=\nu \omega_0$ to obtain,
	\begin{eqnarray}
		E_\text{gw} = \cfrac{\kappa^2}{40}\,\,\mu^2\, \int d\omega^\prime \omega_0^4\,\nu^4\,a^4\,\pi\,\Big[\,\underbrace{f_1(\nu,e)\,\left(H_{\iota \nu}^{(1)}(\iota e\nu)\right)^2 + f_2(\nu,e)\,\left({H_{\iota \nu}^{(1)}}^\prime(\iota e\nu)\right)^2 }_{f(\nu,e)}\,\Big], 
	\end{eqnarray}
\noindent where  $f_{1,2}(\nu,e)$ are given in Eq.~\eqref{eq:f-nu-e}. For a two body scattering with distinct asymptotic states the expression for energy radiated has an extra symmetry factor of $(1/2)$ compared to  Eq.~\eqref{dEdt-2}. This is to compensate for the overcounting for the two body scattering with distinct asymptotic states which needs to be taken into account for the hyperbolic orbit. For the periodic elliptical orbits this symmetry factor is not there.  

Thus the spectrum of energy radiated is given by
	\begin{eqnarray}\label{eq:hyperbolic-power}
		P(\omega^\prime)  = \cfrac{ d E_\text{gw}}{d\omega^\prime} = \cfrac{\kappa^2}{40}\,\,\mu^2\,\omega_0^4\,\nu^4\,a^4\,\pi\,\,f(\nu,e)\,, 
	\end{eqnarray}
Thus, in our approach where we describe the rate of emission through Fermi’s golden rule, and obtain the power spectrum in terms of products of stress-energy tensor components, we reproduce the same expression for $P(\nu)$ as that obtained using the quadrupole formula \cite{Garcia-Bellido:2017knh}. 
 
\begin{figure}[h]
	\includegraphics[scale=0.85]{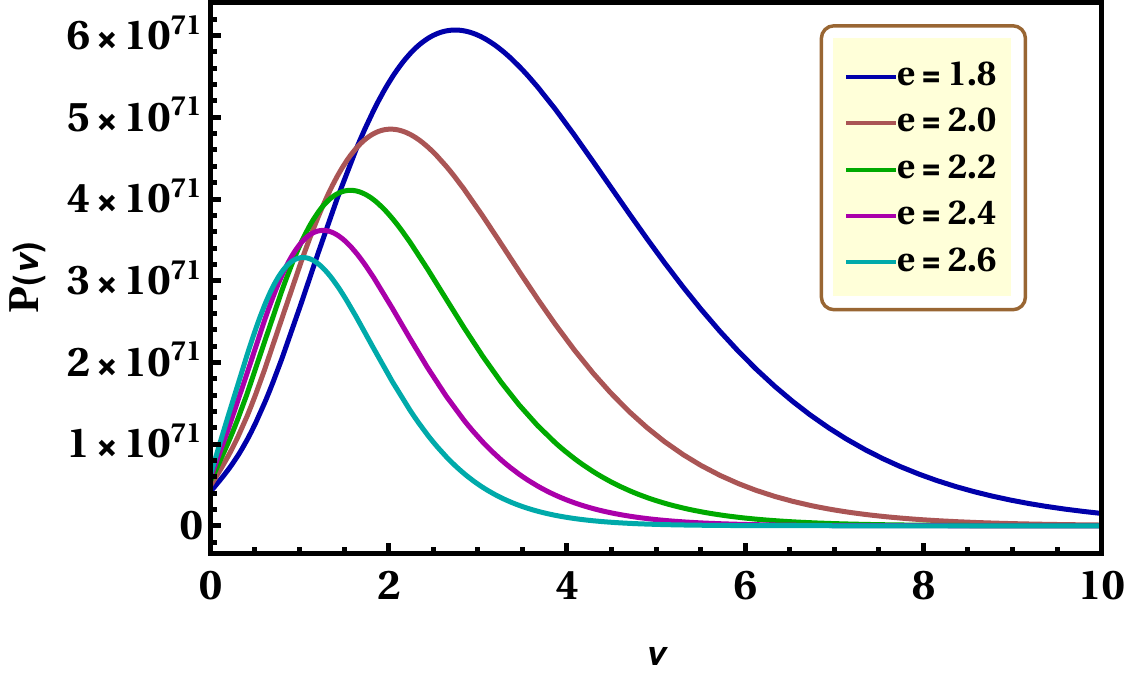}
	\caption{The power spectrum given in Eq.~\eqref{eq:hyperbolic-power} , in frequency space, of gravitational wave radiation from binaries in  hyperbolic orbits with varying eccentricities. For this plot we have taken the following parameters $m_1 = m_2 = 30\,M_\odot,\,\, a = 0.01\,\text{AU}$. The power spectrum is non-zero at zero frequency which represents the energy radiated as the memory signal.   }
	\label{fig:power-spectrum}
\end{figure}
 We have highlighted the features of $P(\nu)$ for different values of eccentricity in Fig.~\ref{fig:power-spectrum}. The total energy loss can be obtained from the power spectrum and is given by:	
	\begin{eqnarray}\label{eq:total-energy-loss}
		\Delta E = \int_{-\infty}^{\infty} dt \,\frac{dE_{\rm gw}}{dt} = \int_{0}^{\infty} d\omega^\prime P(\omega^\prime).
	\end{eqnarray}
	Explicit integration over products of Hankel functions is difficult, especially when the order $p$ of the function $H^{(1)}_p (q)$ also depends on the integration variable. To discern the behaviour of the integral one can do a numerical evaluation of Eq.~\eqref{eq:total-energy-loss}. Upon integrating the power spectrum $P(\nu)$ for $\nu \in [0, \nu_\text{max}]$ and selecting larger and larger values for $\nu_\text{max}$ successively, the result is found to converge to a constant value. This is expected behaviour based on the features of the plot in Fig.~\ref{fig:power-spectrum}, where the curve for each choice of eccentricity flattens out at small enough values of frequency.
\section{Linear memory from hyperbolic orbits}\label{sec:linear-memory}
\subsection{Power emitted at zero frequency}
\noindent Computing the zero frequency limits of the stress-tensors is equivalent to obtaining approximations for them  in the \\$\nu$ $\to$ 0 limit. As $\nu e$ $\to$ 0, the Hankel function and its first derivative assume the following form:
\begin{eqnarray}\label{eq:Hankel-1-0-lim}
	H_{\iota \nu}^{(1)}(\iota e\nu) \simeq \cfrac{2\iota}{\pi} \ln (\nu e), \hspace{1.2cm}
	H_{\iota \nu}^{(1)\prime}(\iota e\nu) \simeq \cfrac{2}{\pi \nu e}.  
\end{eqnarray}
Substituting the above in the expressions for $T_{xx}$, $T_{yy}$ and $T_{xy}$, in Eqs.~\eqref{eq:Txx-hyperb} and \eqref{eq:Tyy-xy-hyperb} yields,
\begin{eqnarray}\label{eq:T_ij-0-lim-hyperp}
	T_{xx} &=& - \cfrac{2\,\mu\, a^{2} \, \omega_0}{e^{2}} \bigg[ \ln(\nu e) + (e^{2} - 1) \bigg], \hspace{0.4cm}
	\hspace*{1.5cm}    T_{yy} = - \cfrac{2\,\mu\, a^{2} \, \omega_0}{e^{2}} \left(e^{2} - 1\right) \bigg[ \ln(\nu e) - 1\bigg], \nonumber\\
	T_{xy} &=& 2\, \iota \, \mu \, \nu\,\omega_0\, a^{2}\, \sqrt{e^{2} - 1} \left[\cfrac{(e^{2} - 1)}{e^{2}} \ln(\nu e) - \cfrac{1}{\nu^{2}\, e^{2}}\right]
	\label{TijHype0}
\end{eqnarray}
Using these, we can rewrite, 
\begin{eqnarray}
	T_{ij}(\nu)\,T^*_{ji}(\nu)-\cfrac{1}{3}\,|\,T^{i}{}_{i}(\nu)\,|^2 
	= 4\, \mu^{2} \, \nu^2\, a^{4} \left[\widetilde{f}_{1}(\nu, e) (\ln(\nu e))^{2}\, + \, \widetilde{f}_{2}(\nu, e) \ln(\nu e)\, +\, \widetilde{f}_{3}(\nu, e) \right].
\end{eqnarray}
with
\begin{eqnarray}
	\widetilde{f}_{1}(\nu, e) &=& \cfrac{2}{e^{4}}(e^{2} - 1)^{3} + \cfrac{6 - 6e^{2} + 2e^{4}}{3\, \nu^{2}\, e^{4}}, \qquad\qquad
	\widetilde{f}_{2}(\nu, e) = \cfrac{2\, (e^{2} - 1)}{\nu^{2}\, e^{4}} - \cfrac{6\, (e^{2} - 1)^{2}}{\nu^{2}\, e^{4}}, \nonumber\\ 
	\widetilde{f}_{3}(\nu, e) &=& \cfrac{2\, (e^{2} - 1)}{\nu^{4}\, e^{4}} +  \cfrac{2\, (e^{2} - 1)^{2}}{\nu^{2}\, e^{4}}.
\end{eqnarray}
The expression for the power spectrum of the gravitational wave, see Eq.~\eqref{eq:hyperbolic-power}, now becomes,
\begin{eqnarray}
	P(\omega^\prime)  = \cfrac{\kappa^{2}}{5}\, \pi\, \mu^{2}\, a^{4}\, \omega_0^{4}\, \nu^{4}\, f(\nu,\, e).
\end{eqnarray}
In the limit $\nu \to 0$,
\begin{eqnarray}
	\lim_{\nu \to 0} \nu^{4}\, f(\nu,\, e) = \cfrac{2\, (e^{2} - 1)}{\pi^{2}\, e^{4}},
\end{eqnarray}
which is finite and different from zero, except for $e = 1$ and $e \rightarrow \infty$. Then the power radiated by the GW of zero frequency is given by,
\begin{eqnarray}
	P(\nu = 0) = \cfrac{32\, G}{5}\,  \mu^{2}\, a^{4}\, \omega_0^{4}\, \cfrac{ (e^{2} - 1)}{e^{4}}.
\end{eqnarray}
  The expression for the power emitted at zero frequency matches the results of refs.~\cite{Garcia-Bellido:2017knh} and \cite{Grobner:2020fnb}. 
\subsection{The memory waveform}
\begin{figure}[h]
\centering
\begin{subfigure}
\centering
	\includegraphics[scale=0.65]{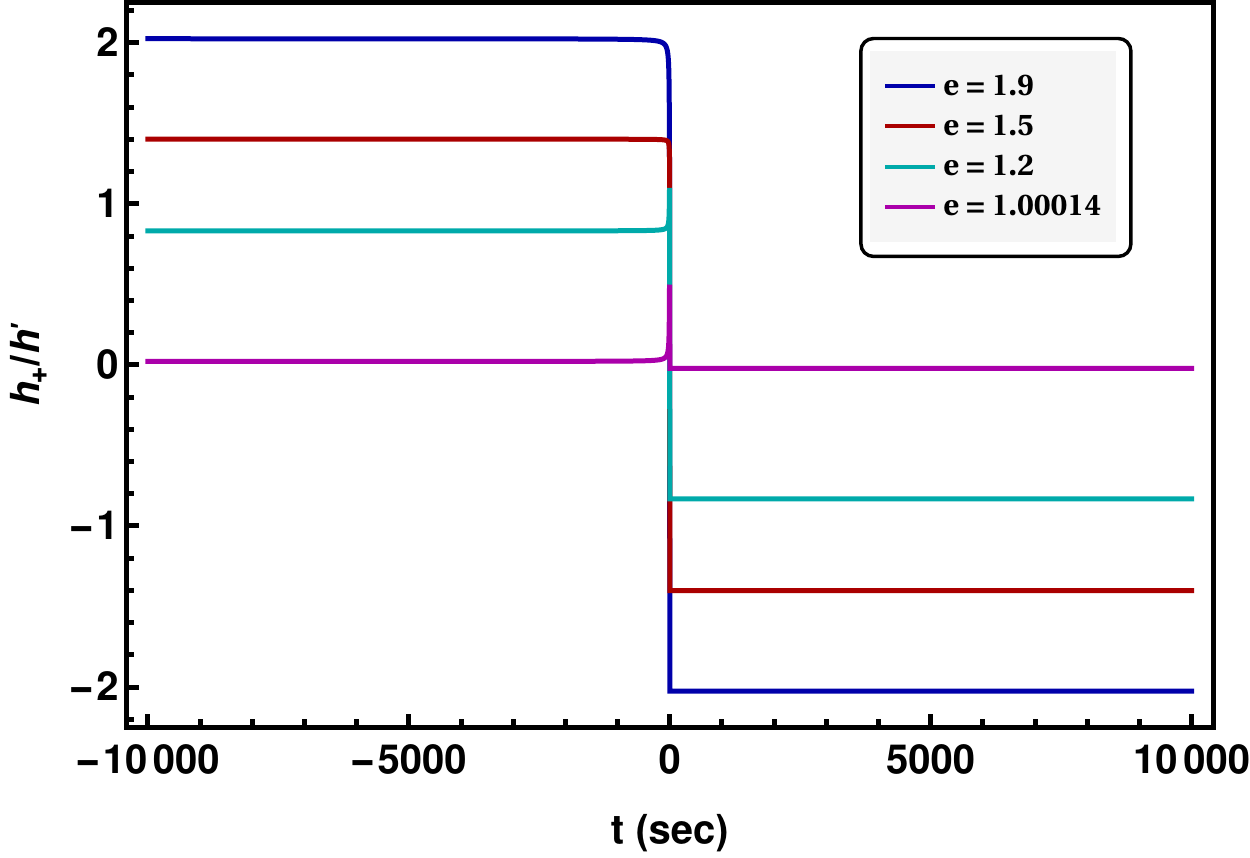}
\end{subfigure}
\hspace*{0.5cm}
\begin{subfigure}
\centering
	\includegraphics[scale=0.65]{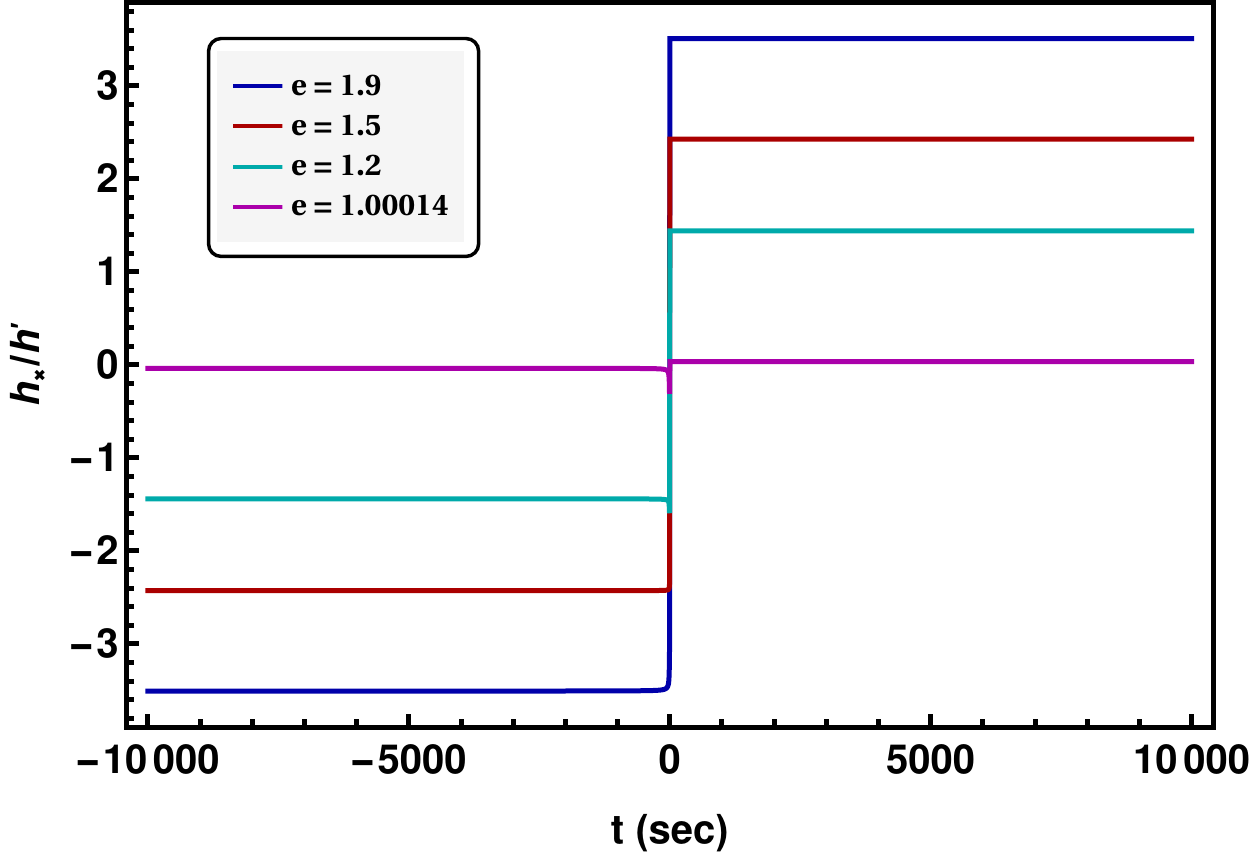}
\end{subfigure}
	\caption{Plots displaying the characteristics of $h_+(t)$ and $h_\times(t)$ for different eccentricities. Here, $\theta$ = 0, $\phi = \frac{\pi}{12}$, and $h^{\prime} = \iota \frac{4G}{r}\left(\frac{2\mu a^{2}\omega_0}{e^{2}}\right)$ is the dimension-ful part of the memory waveform.}
	\label{fig:h-mem-hyperbolic}
\end{figure}

The gravitational wave amplitude corresponding to polarisation $\lambda$, measured by a detector located at a distance $r$, is given (in the frequency domain) by
\begin{eqnarray}
h_\lambda(\omega^\prime, r)= \frac{4 G}{r}  \epsilon^{ij}_\lambda (\vec n) \, T_{ij}(n, \omega^\prime)
\end{eqnarray}
The waveforms for the $+$ and $\times$ polarisations are
\begin{eqnarray}
h_+ (\omega^\prime, r)&=& \iota\cfrac{4G}{r} \epsilon^{ij}_+(\vec n) T_{ij}(\vec n, \omega^\prime)= \iota\cfrac{4G}{r} \left(\vec{e}_{\theta i}  \vec{e}_{\theta j} -\vec{e}_{\phi i} \vec{e}_{\phi j} \right) T_{ij}(\vec n, \omega^\prime)\nn\\
&=&  \iota \frac{4G}{r} \Big( T_{xx}(\cos^2 \phi -\sin^2\phi \cos^2 \theta) +T_{yy} (\sin^2 \phi -\cos^2\phi \cos^2 \theta) - T_{zz}\sin^2 \theta \nn\\
&& \qquad -T_{xy} \sin 2\phi (1+\cos^2\theta) +T_{xz} \sin\phi \sin 2\theta +T_{yz} \cos\phi \sin 2 \theta \Big), \nn\\
&&\nn\\
h_\times(\omega^\prime,r)&=& \iota\frac{4G}{r} \epsilon^{ij}_\times(\vec n) T_{ij}(\vec n, \omega^\prime)= \iota\frac{4G}{r} \left( \vec{e}_{\theta i}  (\vec {e}_\phi)_j + (\vec {e}_\phi)_i (\vec {e}_\theta)_j \right) T_{ij}(\vec n, \omega^\prime)\nn\\
&=& \iota \frac{4G}{r} \Big( (T_{xx}-T_{yy}) \sin 2 \phi \cos \theta+ 2 T_{xy} \cos 2\phi\cos \theta) - 2 T_{xz} \cos\phi \sin \theta +2 T_{yz} \sin\phi \sin  \theta \Big)\,.
\label{Across-hype}
\end{eqnarray}
We have used the spherical coordinates to describe the polarisation of a GW traversing in the radial $\vec n$ direction.  The time domain waveforms can be obtained after evaluating the Fourier transforms of $h_{+} (\omega^\prime, r)$ and $h_{\times} (\omega^\prime, r)$. This can be accomplished numerically and the results have been presented in Fig.~\ref{fig:h-mem-hyperbolic}

\section{Comparison with memory waveforms constructed using soft theorems}\label{sec:soft-theorem-comparison}

\begin{figure}[h]
	\includegraphics[scale=0.5]{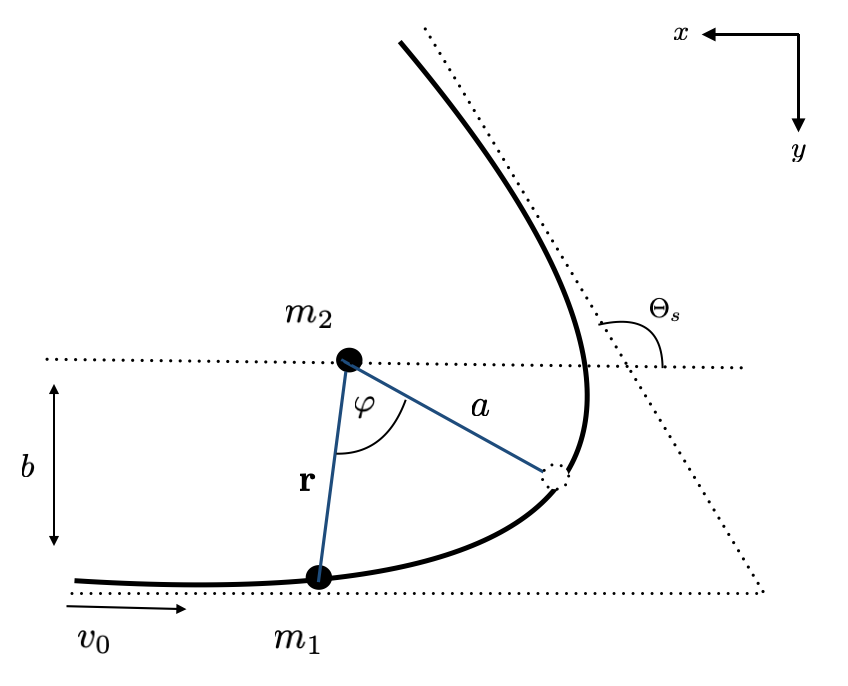}
	\caption{Schematic representation of a hyperbolic encounter between two black holes of masses $m_1$ and $m_2$ ($m_1 < m_2$), in the rest frame of the heavier body. Here, $v_0$ denotes the asymptotic incoming velocity of the lighter body, $b$ and $\Theta_s$ correspond to the impact parameter and the scattering angle and $r_m$ represents the distance of the closest approach.}
	\label{fig:orbit}
\end{figure}

\noindent According to the soft-graviton theorems, the amplitude of a graviton emission from the scattering of $n$-particles of  momenta $p_a$ can be written as the product of a kinematical factor and the amplitude ${\cal A}_{n} (p_a)$ of the n-particle scattering as \cite{Cachazo:2014fwa,Sahoo:2018lxl},
\begin{eqnarray}
{\cal A}_{n+1}(p_a,q)=  \frac{\kappa}{2}  \epsilon^*_{\lambda \mu \nu} \sum_{a=1}^{n} \Big( \frac{  p_{a\mu}  p_{a \nu}}{  p_a\cdot q}  -\iota \frac{   p_{a}^{\mu}  q_{ \beta} J_a^{\beta \nu}}{  p_a\cdot q}   - \frac{  q_\alpha  q_{ \beta} J_a^{\alpha \mu}J_a^{\beta \nu}}{  p_a\cdot q} \Big) {\cal A}_{n} (p_a)
\label{ANLO}.
\end{eqnarray}
Here, $J_a^{\alpha \beta}=x_a^\alpha\, p_a^\beta-x_a^\beta\, p_a^\alpha + S_a^{\alpha \beta}$ describes the total angular momentum of particle $`	`a"$. The series of the soft factors are at the same order in the gravitational coupling but in increasing powers of the graviton frequency $q_0=\omega$. The leading term goes as $ \omega^{-1}$ while the sub-leading terms go as $\sim \omega^0$ and $\sim \omega^1$ respectively. The gravitational waveform at distance $r$ of the radiated soft-graviton of momentum $q$ is given by
\begin{eqnarray}
h_{\mu \nu}(r,q)= \frac{4G}{r} \Big( \frac{  p_{a\mu}  p_{a \nu}}{  p_a\cdot q}  -\iota \frac{   p_{a}^{\mu}  q_{ \beta} J_a^{\beta \nu}}{  p_a\cdot q}   - \frac{  q_\alpha  q_{ \beta} J_a^{\alpha \mu}J_a^{\beta \nu}}{  p_a\cdot q} \Big).
\end{eqnarray}
There are logarithmic corrections to the leading order terms suppressed by $G$ which give a tail contribution  to the linear memory signal (which as a function of time goes as $1/t$), \cite{Laddha:2018myi, Sahoo:2018lxl, Laddha:2018vbn, Saha:2019tub}. The low frequency graviton signal from a generic hard scattering can be written as \cite{Laddha:2018myi},
\begin{eqnarray}\label{softAB}
h_{ij}(\omega)= \iota \omega^{-1} A_{ij} + B_{ij} \ln \omega^{-1} +\cdots ,
\end{eqnarray}
where the coefficients $A_{ij}$ and $B_{ij}$ can be obtained in terms of incoming and outgoing momenta \cite{Laddha:2018myi, Sahoo:2018lxl}. 
Using the relations $\omega=\nu \omega_0$ with $\omega_0=(GM/a^3)^{1/2}$, where $M$ and $a$ are  the total mass and the semi-major axis of the hyperbolic orbit respectively, the memory waveforms obtained in  Eqs.~\eqref{eq:T_ij-0-lim-hyperp} can be rewritten as
\begin{eqnarray}\label{eq:h_ij-0-lim-hyperp}
	h_{xx} &=& -  \frac{4G}{r} \cfrac{2\,\mu\, a^{2} \, \omega_0}{e^{2}} \bigg[ \ln\left(\frac{\omega e}{\omega_0}\right) + (e^{2} - 1) \bigg], \nn\\
	h_{yy} &=& - \frac{4G}{r}  \cfrac{2\,\mu\, a^{2} \, \omega_0}{e^{2}} \left(e^{2} - 1\right) \bigg[ \ln\left(\frac{\omega e}{\omega_0}\right) - 1\bigg], \nonumber\\
	h_{xy} &=&  \frac{4G}{r} \,\frac{2\, \iota \, \mu \, a^{2}\,\omega_0}{e^2}\, \sqrt{e^{2} - 1} \left[\cfrac{(e^{2} - 1) \omega}{ \omega_0} \ln\left(\frac{\omega e}{\omega_0}\right) - \cfrac{\omega_0}{\omega\, }\right]\,.
\end{eqnarray}
From the above equation, we see that the memory signal in frequency space has both the $1/\omega$ and $\ln \omega$ terms as predicted from the general soft-graviton amplitude calculation, see  Eq.~\eqref{softAB}. To compare the coefficients $A_{ij}$ and $B_{ij}$ of the general result \cite{Laddha:2018myi} with the particular case of hyperbolic orbits we express the memory waveform components in Eq.~\eqref{eq:h_ij-0-lim-hyperp} in terms of the initial velocity $\vec v$, ($|\vec v| = v_0$) and impact parameter $b$ as shown in Fig.~\ref{fig:orbit}.

\noindent In the Keplerian orbit there are two conserved quantities $E= \frac{1}{2}\mu v^2$ and $L=\mu b v$ (where $\vec v$ is the velocity of the reduced mass). The semi-major axis $a$ is  given by $a=GM/v_{0}^2$. The impact parameter $b$ is related to the eccentricity as $e=(1+b^2/a^2)^{1/2}= (1+L^2/(\mu v_{0} a)^2))^{1/2}$. Doing an expansion of $h_{ij}$ in powers of the angular momentum $L$ and retaining terms to the leading order in $L$, the coefficient of the $\iota \omega^{-1}$ term follows from Eq.~\eqref{eq:h_ij-0-lim-hyperp} as
\be\label{hype-Aij}
A_{xy}= - \cfrac{4G}{r}  2\mu v_{0}^2\, \cfrac{L v_{0}}{\mu G M} \left(1-\left(\frac{L v_{0}}{\mu G M} \right)^2\right) , \quad A_{xx}=0\,, \quad A_{yy}=0\,.
\ee
and the coefficient of the $\ln\omega^{-1}$ terms in the memory signal  are
\bea\label{hype-Bij}
B_{xx}= \frac{4G}{r}2 \mu \frac{ G M}{v_{0}} \left(1-\left(\frac{L v_{0}}{\mu G M} \right)^2\right), \quad B_{yy}=\frac{4G}{r}2 \mu \frac{ G M}{v_{0}} \left(\frac{L v_{0}}{\mu G M} \right)^2\,, \quad
B_{xy}= 0.
\eea

\noindent The coefficient of $\ln \omega$ in the general expressions for the energy-momentum tensor components $T_{ij}$ outlined in \cite{Saha:2019tub}, based on soft-graviton theorems, assume the same form as the contents of Eqs.~\eqref{eq:T_ij-0-lim-hyperp} for the specific case of a hyperbolic encounter. Based on the parametrizations for spatial and temporal coordinates, i.e., 
\begin{eqnarray}
	x(\xi) = a\,(e - \cosh\xi),\qquad y(\xi) = b\,\sinh\xi, \qquad z(\xi) = 0, \qquad \cfrac{\omega^\prime}{\nu}\,t = \omega_0\, t = (e\sinh\xi - \xi),
\end{eqnarray}
we can compute the components of the initial and final velocities of the reduced mass as
\begin{eqnarray}
	v^{in}_i = \cfrac{dx_i}{dt}\,\Bigg|_{t=-\infty} = \bigg(\cfrac{dx_i(\xi)}{d\xi}\bigg)\bigg/\bigg(\cfrac{dt(\xi)}{d\xi}\bigg)\,\Bigg|_{\xi=-\infty} \hspace{0.6cm}\text{and}\hspace{0.6cm} v^{out}_i = \cfrac{dx_i}{dt}\,\Bigg|_{t=\infty} = \bigg(\cfrac{dx_i(\xi)}{d\xi}\bigg)\bigg/\bigg(\cfrac{dt(\xi)}{d\xi}\bigg)\,\Bigg|_{\xi=\infty}.
\end{eqnarray}
Using $v_0 = \omega_0\,a$ and $b = a\,\sqrt{e^2 - 1}$, we obtain, for the co-ordinate system of Fig.~\ref{fig:orbit},
\begin{eqnarray}
	-v_{x}^{in} = v_{x}^{out}= \frac{v_{0}}{e}, \qquad v_{y}^{in} = v_{y}^{out} = -\frac{v_{0}}{e}\sqrt{e^{2}-1}, \qquad v_{z}^{in} = v_{z}^{out} = 0.
\end{eqnarray}
Substituting these expressions in the general result for the $\ln\omega$ proportional part of the energy-momentum tensor given in \cite{Saha:2019tub}, 
\begin{eqnarray}\label{eq:soft}
	\hat{T}^{X\mu\nu}(k) & = & 2G\, \frac{\ln \{L(\omega - \iota \epsilon)\}}{\{(p_{1}^{in}\cdot p_{2}^{in})^{2}-(p_{1}^{in})^{2}(p_{2}^{in})^{2}\}^{3/2}}\Bigr[\Bigr\{ \frac{n\cdot p_{2}^{in}}{n\cdot p_{1}^{in}} (p_{1}^{in})^{\mu}(p_{1}^{in})^{\nu} + \frac{n\cdot p_{1}^{in}}{n\cdot p_{2}^{in}} (p_{2}^{in})^{\mu}(p_{2}^{in})^{\nu} \Bigr\} \bigr(p_{1}^{in}\cdot p_{2}^{in}\bigr) \nonumber \\
	&& \times \Bigr\{\frac{3}{2}(p_{1}^{in})^{2}(p_{2}^{in})^{2} -(p_{1}^{in}\cdot p_{2}^{in})^{2} \Bigr\} + \frac{1}{2}(p_{1}^{in})^{2}(p_{2}^{in})^{2}\{ (p_{2}^{in})^{2}(p_{1}^{in})^{\mu}(p_{1}^{in})^{\nu} + (p_{1}^{in})^{2}(p_{2}^{in})^{\mu}(p_{2}^{in})^{\nu}\}   \nonumber \\
	&& - 2 \{ (p_{1}^{in})^{\mu} (p_{2}^{in})^{\nu} + (p_{1}^{in})^{\nu} (p_{2}^{in})^{\mu}\}(p_{1}^{in}\cdot p_{2}^{in}) \Bigr\{\frac{3}{2}(p_{1}^{in})^{2}(p_{2}^{in})^{2} -(p_{1}^{in}\cdot p_{2}^{in})^{2} \Bigr\} \Bigr] \nonumber \\
	&& + 2G\, \frac{\ln \{L(\omega + \iota \epsilon)\}}{\{(p_{1}^{out}\cdot p_{2}^{out})^{2}-(p_{1}^{out})^{2}(p_{2}^{out})^{2}\}^{3/2}}\Bigr[\Bigr\{ \frac{n\cdot p_{2}^{out}}{n\cdot p_{1}^{out}} (p_{1}^{out})^{\mu}(p_{1}^{out})^{\nu} + \frac{n\cdot p_{1}^{out}}{n\cdot p_{2}^{out}} (p_{2}^{out})^{\mu}(p_{2}^{out})^{\nu} \Bigr\} \bigr(p_{1}^{out}\cdot p_{2}^{out}\bigr) \nonumber\\
	&& \times \Bigr\{\frac{3}{2}(p_{1}^{out})^{2}(p_{2}^{out})^{2} -(p_{1}^{out}\cdot p_{2}^{out})^{2} \Bigr\} + \frac{1}{2}(p_{1}^{out})^{2}(p_{2}^{out})^{2}\{ (p_{2}^{out})^{2}(p_{1}^{out})^{\mu}(p_{1}^{out})^{\nu} + (p_{1}^{out})^{2}(p_{2}^{out})^{\mu}(p_{2}^{out})^{\nu}\}  \nonumber \\
	&& - 2 \{ (p_{1}^{out})^{\mu} (p_{2}^{out})^{\nu} + (p_{1}^{out})^{\nu} (p_{2}^{out})^{\mu}\}(p_{1}^{out}\cdot p_{2}^{out}) \Bigr\{\frac{3}{2}(p_{1}^{out})^{2}(p_{2}^{out})^{2} -(p_{1}^{out}\cdot p_{2}^{out})^{2} \Bigr\} \Bigr] \\\nonumber
\end{eqnarray}
we get
\begin{eqnarray}\label{eq:T_ij-soft-thm}
	T_{xx} = - \frac{Gm_{1}m_{2}}{v_{0}} \frac{2}{e^{2}}\ln\omega + \mathcal{O}(v_{0}), \hspace{1cm}
	T_{yy} = - \frac{Gm_{1}m_{2}}{v_{0}} 2 \left( 1 - \frac{1}{e^{2}}\right) \ln\omega + \mathcal{O}(v_{0}), \hspace{1cm} T_{xy} = 0.
\end{eqnarray}
Expressing the velocity and the fundamental frequency in terms of the semi-major axis,
\begin{eqnarray}
	v_0 = \sqrt{\cfrac{G(m_1 + m_2)}{a}} \hspace{1cm}\text{and}\hspace{1cm} \omega_0 = \sqrt{\cfrac{G(m_1 + m_2)}{a^3}}.
\end{eqnarray}
allows us to identify $Gm_1m_2 / v_0 = \mu a^2 \omega_0$ and this establishes the equivalence between Eq.~\eqref{eq:T_ij-soft-thm} and Eq.~\eqref{eq:T_ij-0-lim-hyperp}. Thus, we note an agreement between our result for the linear memory waveform and the waveform constructed using soft theorems \cite{Laddha:2018myi, Sahoo:2018lxl, Laddha:2018vbn, Saha:2019tub}.
 
\section{Gravitational radiation from binaries in elliptical orbits}\label{sec:elliptical} 
	
\noindent For eccentric elliptical orbits, Peters and Mathews \cite{Peters-1,Peters-2} calculated
the average energy and angular momentum emission rates at
Newtonian order. Their calculation has been improved to the 3-PN level \cite{Blanchet:2004ek,Arun:2007sg}, including nonlinear tail-effect (which arises from the scattering of gravitational waves by the near-field potential) at 3-PN \cite{Arun:2007rg}. In this section  we compute the frequency spectrum of the energy radiated following \cite{Mohanty:1994yi,KumarPoddar:2019jxe,KumarPoddar:2019ceq,KumarPoddar:2019ceq,Poddar:2021yjd}. The radiated energy acts as the source term for secondary gravitational waves which carry the non-linear memory signal. 

\noindent To describe a compact binary system, comprised of stars having masses $m_1$ and $m_2$, in an elliptical orbit, the following quantities are of relevance:
	\begin{itemize}
		\item Motion around the common centre of mass can be described in terms of the reduced mass $\mu = \cfrac{m_1\,m_2}{m_1+m_2}$ and the total mass $M = (m_1 + m_2)$.
		\item The coordinates of the elliptical Keplerian orbit can be parametrized as:
		\begin{eqnarray}\label{elliptic cordinate}
			\quad x(\xi) = a\,(\cos\xi - e),\qquad y(\xi) = b\,\sin\xi, \qquad z(\xi) = 0, \qquad \cfrac{\omega_n^\prime}{n}\,t = \omega_0\, t = (\xi - e\sin\xi).
		\end{eqnarray}	
		Here, $a$ and $b \,\,(= a \sqrt{1 - e^2},\,\, e < 1)$ denote the semi-major and semi-minor axes respectively. The variables $e$ and $\xi \in (0, 2\pi)$ refer to the eccentricity of the orbit and the eccentric anomaly respectively.
				
		\item The angular frequency corresponding to the $n^{th}$ harmonic is denoted as $\omega_{n}^\prime = n\,\omega_0$, with $n = \{0,1,2,....\}$ being a non-negative integer and the fundamental frequency $\omega_0$ can be related to the semi-major axis and the total mass of the system as: $\quad \omega_0 = \left(\cfrac{G\, (m_1 + m_2)}{a^3}\right)^{1/2}$.
	\end{itemize}
The first step in the computation of the rate of energy radiated involves the evaluation of the stress-tensor components in frequency space. Once again starting with the $xx$ component, the calculation proceeds as follows:
\begin{eqnarray}
T_{xx}(\omega_{n}^{\prime}) &=& \cfrac{\mu}{T}\, \int_{0}^{T}\, dt\, \dot{x}^{2}(t) e^{\iota \omega_{n}^{\prime} t} 
= \cfrac{-\iota\,\mu\, \omega_{n}^{\prime}}{T}\,\int_{0}^{T}\, dt\, \dot{x}(t)\, x(t)\, e^{\iota \omega_{n}^{\prime} t}.
\end{eqnarray}
In the above, we have used integration by parts. Next, the parametric form of the orbit coordinates, shown in Eq.~\eqref{elliptic cordinate}, allows us to write,
\begin{eqnarray}
\dot{x}\,dt = \cfrac{dx}{d\xi} \,d\xi = - a \sin \xi\, d\xi.
\end{eqnarray}
Using the above transformation and also substituting for $x$ and $\omega_n^\prime\,t$ in terms of functions of $\xi$, we get
\begin{eqnarray}\label{Txx:ellip}
T_{xx}(\omega_{n}^{\prime}) &=& \cfrac{\iota\, \mu a^{2} \omega_{n}^{\prime 2}}{2 \pi n} \int_{0}^{2\pi} d\xi \,\sin\xi \,(\cos\xi - e)\, e^{\iota n (\xi - e \sin\xi)} 
 = - \cfrac{ \mu a^{2} \omega_{n}^{\prime 2}}{n}\left[\left(\cfrac{1 - e^{2}}{e}\right)J_{n}^{\prime}(ne) - \cfrac{1}{n e^{2}}J_{n}(ne) \right].
\end{eqnarray}
To arrive at the second line of Eq.~\eqref{Txx:ellip}, we replaced the trigonometric functions by the corresponding exponential functions and identified the integral form of Bessel functions of first kind, 
\begin{eqnarray}\label{eq:bessel-fn}
J_n(z) = \cfrac{1}{2\pi}\,\int\limits_{0}^{2\pi}\,e^{\iota n (\xi - e \sin\xi)} d\xi.
\end{eqnarray}
The final expression in terms of $J_n(ne)$ and $J^\prime_n(ne)$ is obtained by utilizing the recurrence relations given below:
\begin{eqnarray}\label{eq:bessel-recurr}
J_{n-1}(z) + J_{n+1}(z) = \cfrac{2\,n}{z}\,J_n(z), \hspace{1cm} J_{n-1}(z) - J_{n+1}(z) = 2\,J^\prime_n(z).
\end{eqnarray}    
The other non-zero components, i.e $T_{yy}$ and $T_{xy}$ can similarly be obtained as:
\begin{eqnarray}\label{Tyy-xy:ellip}
T_{yy}(\omega_{n}^{\prime}) &=& \cfrac{\mu}{T} \int_{0}^{T} dt\, \dot{y}^{2}(t) e^{\iota \omega_{n}^{\prime} t} 
 = \cfrac{ \mu\, \omega_{n}^{\prime 2}\, a^{2}(1 - e^{2})}{n}\left[\cfrac{1}{e} J_{n}^{\prime}(ne) - \cfrac{1}{n e^{2}}J_{n}(ne) \right], \nonumber\\
T_{xy}(\omega_{n}^{\prime}) &=& \cfrac{\mu}{T}\, \int_{0}^{T}\, dt\, \dot{x}(t)\dot{y}(t) e^{\iota \omega_{n}^{\prime} t} 
= \cfrac{\iota \mu\,  \omega_{n}^{\prime 2}\, a^{2}\sqrt{(1 - e^{2})}}{n}\left[-\left(\cfrac{1-e^{2}}{e^{2}}\right)J_{n}(ne)  + \cfrac{1}{n e}J_{n}^{\prime}(ne) \right]. 
\end{eqnarray}
Substituting for $T_{xx}$, $T_{yy}$ and $T_{xy}$ using Eqs.~\eqref{Txx:ellip} and \eqref{Tyy-xy:ellip} gives,
	\begin{eqnarray}
		T_{ij}(\omega^\prime)\,T^*_{ji}(\omega^\prime)-\cfrac{1}{3}\,|\,T^{i}{}_{i}(\omega^\prime)\,|^2 = g(n, e).
	\end{eqnarray}
	where we have defined $g(n,e)$ as,
	\begin{eqnarray}
	g(n,e) &=& J_{n}(ne)^{2}\left[\cfrac{2n^{2}}{e^{4}}(1-e^{2})^{3} + \cfrac{6 - 6e^{2} + 2e^{4}}{3e^{4}} \right] + J_{n}^{\prime}(ne)^{2}\left[\cfrac{2n^{2}}{e^{2}}(1-e^{2})^{2} + \cfrac{2(1 - e^{2})}{e^{2}} \right]  \nonumber\\
	&& +\, J_{n}(ne)J_{n}^{\prime}(ne) \left[\cfrac{(-8 + 14 e^{2} - 6 e^{4})n}{e^{3}} \right].
	\end{eqnarray}
	In this case, the energy loss due to gravitational radiation as given in Eq.~\eqref{dEdt-2} becomes, 
	\begin{eqnarray}
	\cfrac{dE_\text{gw}}{dt}(e) &=&  \cfrac{\kappa^{2}}{8(2\pi)^{2}} \sum_{n = 0} \cfrac{8\pi}{5}\left[T_{ij}(\omega_{n}^{\prime})T_{ji}^{*}(\omega_{n}^{\prime}) - \cfrac{1}{3}|T^{i}_{i}(\omega_{n}^{\prime})|^2\right] \omega_{n}^{\prime 2}
	 =  \cfrac{32\,G}{20}\,\omega_0^{6}\mu^{2}a^{4}\sum_{n = 0}^{\infty} n^{2}g(n,e).
	\end{eqnarray}
The series sum over products of Bessel functions and their derivative, weighted by powers of $n$, can be expressed completely as functions of $e$, using the identities derived in the appendix of \cite{Peters-1} and reproduced in Eq.~\eqref{eq:Peter-Matthews-sum}. This enables the identification of the eccentricity dependent part of the energy loss as	
	\begin{eqnarray}
	\widetilde{g}(e) = \sum_{n = 0}^{\infty} n^{2}\,g(n,e) = \cfrac{4}{(1-e^{2})^{7/2}}\left(1 + \cfrac{73}{24}\,e^{2} + \cfrac{37}{96}\,e^{4} \right).
	\end{eqnarray}
Therefore, the energy radiated as gravitational waves from the binary can be written as:~\cite{Poddar:2021yjd, Mohanty:2022abo}
	\begin{eqnarray}
	\cfrac{dE_\text{gw}}{dt} = \cfrac{32\,G}{5}\,\omega_0^{6}\left(\cfrac{m_{1}m_{2}}{m_{1}+m_{2}}\right)^{2}a^{4}\cfrac{1}{(1 - e^{2})^{7/2}}\left(1 + \cfrac{73}{24}\,e^{2} + \cfrac{37}{96}\, e^{4} \right).
	\end{eqnarray}
Once again, we note an agreement between the expression for rate of energy loss computed using our approach and the expression obtained using the quadrupole formula~\cite{Peters-1, Peters:1964zz}. 
	
\section{Non-linear memory from elliptical orbits}\label{sec:ellip-non-lin-mem}
The rate of energy radiated with respect to time and solid angle is described using the following formula:
\begin{eqnarray}
	\cfrac{dE_\text{gw}}{dt^\prime \, d\Omega^\prime} &=&  \cfrac{\kappa^{2}}{8(2\pi)^{2}} \sum_{n = 0} \bigg[T_{ij}(\omega_n^\prime)\,T^*_{kl}(\omega_n^\prime)\,\Lambda_{ij,kl}(\hat n^\prime)\bigg]\, \omega_{n}^{\prime 2}\,.
\end{eqnarray}
The stress-energy tensor components  $T_{ij}$, where $i,j = x,y,z$, can be collected together in matrix form as: 
\begin{eqnarray}
	\textsf{T}\, (n,e) = \mu\, a^2\, \omega_0^2\,\begin{pmatrix}
		q_1(n,e) && \iota\,q_2(n,e) && 0\\
		\iota\,q_2(n,e) && q_3(n,e) && 0\\
		0 && 0 && 0
	\end{pmatrix}.
\end{eqnarray}
The non-zero elements of the matrix are functions of $n$, corresponding to the $n$th harmonic, and $e$, the eccentricity of the orbit and these are given as 
\begin{eqnarray}
	q_1(n,e) &=& -\cfrac{n\,\left(1 - e^{2}\right)}{e}\,J_{n}^{\prime}(ne) + \cfrac{1}{e^{2}}\,J_{n}(ne),  \nonumber\\
	q_2(n,e) &=& \cfrac{\left(1 - e^{2}\right)^{1/2}}{e}\,J_{n}^{\prime}(ne) - \cfrac{n\,\left(1 - e^{2}\right)^{3/2}}{e^{2}}\,J_{n}(ne), \nonumber\\
	q_3(n,e) &=& \cfrac{n\,\left(1 - e^{2}\right)}{e}\,J_{n}^{\prime}(ne) - \cfrac{\left(1 - e^{2}\right)}{e^{2}}\,J_{n}(ne).
\end{eqnarray}	

\begin{figure}[h]
	\includegraphics[scale=0.35]{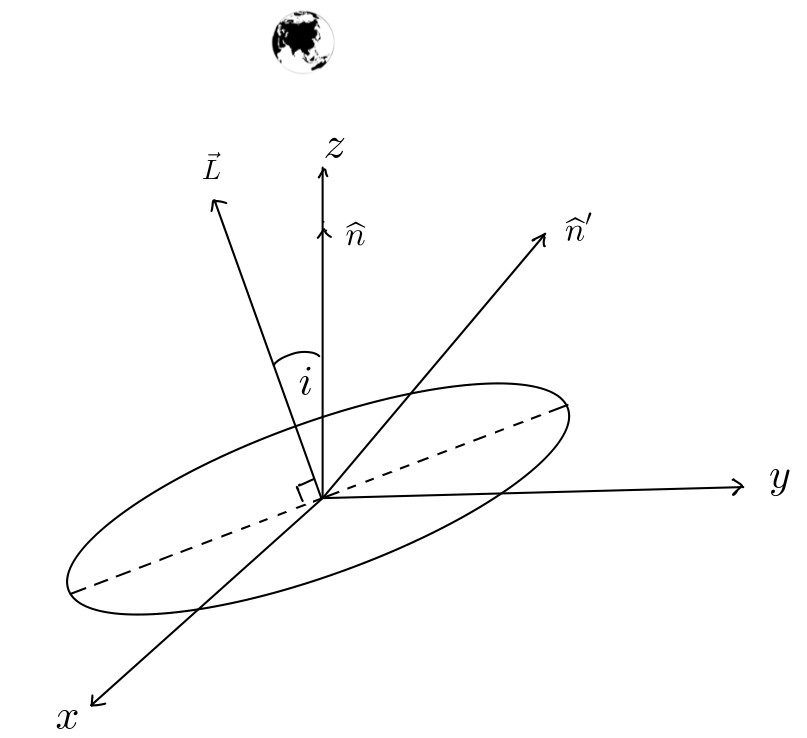}
	\caption{ The axis of rotation of the binary designated by $\vec{L}$ makes an angle $i$ with the $z$-axis and therefore lies in the $y$-$z$ plane. The primary graviton emits a secondary graviton at $\vec r^\prime= r^\prime \hat n^\prime= r^\prime (\sin \theta^\prime \cos\phi^\prime, \sin\theta^\prime \sin \phi^\prime, \cos\theta^\prime)$ and the secondary graviton travels to the earth located at $\vec r= r \hat n=(0,0,1)$.}
	\label{fig:Non-Linear-Memory}
\end{figure}
\noindent We follow the orientation for the system as shown in Fig.~\ref{fig:Non-Linear-Memory}, where the axis of rotation of the binary system lies in the $y$-$z$ plane and is counter-clockwise rotated, by an angle $i$, with respect to the $z$ axis, i.e. $\vec{L} = (0,-\sin i, \cos i)$. The stress-energy matrix for the rotated system is given as $\textsf{T}^\prime = \mathcal{R}\,\textsf{T}\,\mathcal{R}^\text{T}$, with $\mathcal{R}$ being the rotation matrix defined in Eq.~\eqref{eq:rotation}. This allows us to write,
\begin{eqnarray}
 \textsf{T}^\prime =\mu\, a^2\, \omega_0^2 \begin{pmatrix}
		q_1(n,e) && \iota \,q_2(n,e)\, \cos i && \iota \,q_2(n,e)\, \sin i\\
		\iota \,q_2(n,e)\, \cos i && q_3(n,e)\, \cos^2 i && q_3(n,e)\, \cos i\sin i\\
		\iota \,q_2(n,e)\, \sin i && q_3(n,e)\, \cos i\sin i && q_3(n,e)\, \sin^2 i
	\end{pmatrix} .
\end{eqnarray}
Expanding the matrix product $T_{ij}\,T^*_{kl}\,\Lambda_{ij,kl}(\hat n^\prime)$ in terms of products of $T_{ij}$ and $n^\prime_i$ yields, 
\begin{eqnarray}\label{eq:tensor-product-index-2}
	T_{ij}\,T^*_{kl}\,\Lambda_{ij,kl}(\hat n^\prime) &=& T_{ij}\,T^*_{ji} - 2\, T_{ij}\,T^*_{jl}\,n_i^\prime\, n_l^\prime  + \cfrac{1}{2}\,T_{ij}\, T^*_{kl}\, n_i^\prime\,  n_j^\prime\,  n_k^\prime\,  n_l^\prime   + \cfrac{1}{2}\,\left( T_{ii}\,T^*_{kl}\, n_k^\prime\, n_l^\prime + T_{ij}\,T^*_{kk}\, n_i^\prime\, n_j^\prime - T_{ii}\,T^*_{kk} \right).
\end{eqnarray}
Since, $T_{ij}$'s are functions of $n$ and $e$, while $\hat{n}^\prime$ is parametrized in terms of the angles $\theta^\prime$, $\phi^\prime$, the matrix product can be expressed as an overall function of $n$, $e$, $\theta^\prime$, and $\phi^\prime$, i.e.,
\begin{eqnarray}\label{eq:tensor-product-index-3}
	T_{ij}(n,e) \,T^*_{kl}(n,e)\,\Lambda_{ij,kl}(\theta^\prime, \phi^\prime) &=&\mu^2 a^4\, \omega_0^4\, \,\mathcal{I}(n,e,\theta^\prime, \phi^\prime)\, ,
\end{eqnarray}
with
\begin{eqnarray}
	 \mathcal{I}(n,e,\theta^\prime, \phi^\prime) &=&
	 \cfrac{1}{2}\,q^2_1(n,e)\left[1-p_1^2(\theta^\prime, \phi^\prime)\right]^2  + \cfrac{1}{2}\,q^2_3(n,e)\left[1-p_2^2(\theta^\prime, \phi^\prime)\right]^2 - 2\,q_1(n,e)\,q_3(n,e) \nonumber\\
	 &&+\, 2\,q^2_2(n,e)\left[1-p_1^2(\theta^\prime, \phi^\prime)\right]\left[1-p_2^2(\theta^\prime, \phi^\prime)\right] + q_1(n,e)\,q_3(n,e)\left[1+p_1^2(\theta^\prime, \phi^\prime)\right]\left[1+p_2^2(\theta^\prime, \phi^\prime)\right].
\end{eqnarray}
Here,
\begin{eqnarray}
	p_1(\theta^\prime, \phi^\prime) = \sin\theta^\prime\,\cos\phi^\prime\, , \quad \text{and} \quad p_2(\theta^\prime, \phi^\prime) = \cos i \sin \theta^\prime \sin \phi^\prime + \sin i \cos \theta^\prime\,.
\end{eqnarray}
Computation of the transverse-traceless wavefunction involves\footnote{See appendix \ref{app:non-lin-mem-gen-deriv} for details on the derivation of the general expression for non-linear memory waveform and appendix \ref{app:non-lin-mem-circular} for the simpler example of a circular orbit.} (i) an angular integral over $\theta^\prime$, $\phi^\prime$ and (ii) a sum over the orders ($n$) of the Bessel functions. This ultimately leads to an eccentricity dependent result:  
\begin{eqnarray}
	\Big[\textsf{A}_{ij}\Big]^\text{TT}(e) = \sum_{n = 0}^{\infty}\, \, \int_{4\pi} d\,\Omega^\prime\,\mathcal{I}(n,e,\theta^\prime, \phi^\prime)\, \cfrac{ \Lambda_{ij,kl}(\hat n)\, n^\prime_k n^\prime_l }{(1-\hat n^\prime \cdot \hat n)}.
\end{eqnarray}
The tensor product within the angular integral can be further expanded as:
\begin{eqnarray}
	\Lambda_{ij,kl}(\hat n)\, n^\prime_k n^\prime_l =  \left[ \left(n^\prime_i n^\prime_j -\cfrac{1}{2}\,\delta_{ij} + \cfrac{1}{2}\,n_i n_j \right) - \left( n_i n^\prime_j + n^\prime_i n_j\right)\left( \hat n^\prime \cdot \hat n  \right) + \cfrac{1}{2}\,\left(\delta_{ij} + n_i n_j\right)\left( \hat n^\prime \cdot \hat n  \right)^2 \right].
\end{eqnarray}
Since, $\hat{n} = (0,0,1)$, $\hat n^\prime \cdot \hat n = \cos \theta^\prime$ and substituting for the components of $\hat{n}$, $\hat{n}^\prime$ and $\delta_{ij}$, the angular integrals can be evaluated, for $i,j = x,y$ as follows:
\begin{eqnarray}\label{Eq:elipA11}
	\Big[\textsf{A}_{xx}\Big]^\text{TT}(e) &=& \sum_{n = 0}^{\infty}\, \, \int_{0}^{\pi} \sin\theta^\prime \,d\theta^\prime\, \int_{0}^{2\pi} d\phi^\prime \, \mathcal{I}(n,e,\theta^\prime, \phi^\prime) \,\times\,\cfrac{1}{2}\left( 1 + \cos \theta^\prime\right)\,\cos 2\phi^\prime\, , \nonumber\\
	&=& \cfrac{2\pi}{15}\,\sum_{n = 0}^{\infty}\, \bigg[ \Big( 6\,q_1(n,e)\,q_3(n,e) -3\,q^2_1(n,e) - 8\,q^2_2(n,e) \Big) \nonumber\\
	&& \qquad\qquad + \Big( 8\,q^2_2(n,e) + 2\,q^2_3(n,e) -6\,q_1(n,e)\,q_3(n,e) \Big) \cos^2 i + q^2_3(n,e) \cos^4 i\bigg]\, \nonumber \\
	& = & \cfrac{2\pi}{15} \left(C_{0}(e) + C_{2}(e)\cos^{2} i + C_{4}(e)\cos^{4} i\right)\, .
\end{eqnarray}
The sum over $n$ can be evaluated, using the identities given in Eq.~\eqref{eq:Peter-Matthews-sum}, to obtain the coefficients $C_{i}(e)$, $i = 0,2,4$:

\begin{eqnarray}
	C_{0}(e) & = &  \sum_{n = 0}^{\infty}\,\bigg[  \cfrac{(6e^{2} - 9)}{e^{4}}\, n^{2} \left[J_{n}(ne)\right]^2 -  \cfrac{8(1-e^{2})^{3}}{e^{4}}\, n^{4} \left[J_{n}(ne)\right]^2 - \cfrac{8(1-e^{2})}{e^{2}}\, n^{2} \left[J^\prime_{n}(ne)\right]^2 - \cfrac{9(1-e^{2})^{2}}{e^{2}}\, n^{4} \left[J^\prime_{n}(ne)\right]^2 \nonumber \\
	&  &  \qquad\qquad + \cfrac{(1-e^{2})}{e^{3}}( 34 - 22e^{2})\, n^{3} \left[J_{n}(ne)\,J^\prime_{n}(ne)\right] \bigg] \nonumber \\
	& = &  - \cfrac{1}{4 e^{2}(1 - e^{2})^{7/2}} \,\bigg[68\, e^{2} + \cfrac{1565}{8}\, e^{4} + \cfrac{1533}{64}\,e^{6} \bigg],
	\nonumber
\end{eqnarray}

\begin{eqnarray}\label{eq:C_024-e}
	C_{2}(e) &=& \sum_{n = 0}^{\infty}\,\bigg[ \cfrac{2(4-5e^{2} + e^{4})}{e^{4}}\, n^{2} \left[J_{n}(ne)\right]^2 +  \cfrac{8(1-e^{2})^{3}}{e^{4}}\, n^{4} \left[J_{n}(ne)\right]^2 + \cfrac{8(1-e^{2})}{e^{2}}\, n^{2} \left[J^\prime_{n}(ne)\right]^2 + \cfrac{8(1-e^{2})^{2}}{e^{2}}\, n^{4} \left[J^\prime_{n}(ne)\right]^2 \nonumber \\ 
	&  &  \qquad\qquad + \cfrac{2(1-e^{2})}{e^{3}}(-16 + 13 e^{2})\, n^{3} \left[J_{n}(ne)\,J^\prime_{n}(ne)\right] \bigg] \nonumber\\
	&=&  \cfrac{1}{4 e^{2}(1 - e^{2})^{7/2}} \bigg[64\, e^{2} +     \cfrac{399}{2}\, e^{4} + \cfrac{101}{4}\, e^{6}  \bigg], \nonumber\\
	&& \nonumber\\
	 C_{4}(e) &=&  \sum_{n = 0}^{\infty}\, \cfrac{(1-e^{2})^{2}}{e^{4}} \bigg[ n^{2} \left[J_{n}(ne)\right]^2 + e^{2} n^{4} \left[J^\prime_{n}(ne)\right]^2 - 2 e\, n^{3} \left[J_{n}(ne)\,J^\prime_{n}(ne)\right] \bigg] \nonumber\\
	& = &
	\cfrac{1}{4e^{2}(1 - e^{2})^{7/2}} \bigg[ 4\, e^{2} + \cfrac{125}{8}\, e^{4} + \cfrac{109}{64}\, e^{6} \bigg].
\end{eqnarray}
It can easily be seen that in the limit $e \to 0$, Eq.~\eqref{Eq:elipA11} reduces to
\begin{eqnarray}
\Big[\textsf{A}_{xx}\Big]^\text{TT}\Big(e \to 0 \Big) = -  \cfrac{2\pi^{2}}{15} \sin^{2}{i}\,(17 + \cos^{2}{i}),
\end{eqnarray}

\noindent which exactly matches with its circular counterpart as obtained in Eq.~\eqref{Eq:CircA11}.\\
The cross-component $\big[\textsf{A}_{xy}\big]^\text{TT}(e)$ can similarly be evaluated as:

\begin{eqnarray}
	\Big[\textsf{A}_{xy}\Big]^\text{TT}(e) &=& \sum_{n = 0}^{\infty}\, \, \int_{0}^{\pi} \sin\theta^\prime \,d\theta^\prime\, \int_{0}^{2\pi} d\phi^\prime \, \mathcal{I}(n,e,\theta^\prime, \phi^\prime) \,\times\,\cfrac{1}{2}\left( 1 + \cos \theta^\prime\right)\,\sin 2\phi^\prime = 0.
\end{eqnarray}

\noindent The other components can be obtained using the above through the relations: $	\big[\textsf{A}_{yy}\big]^\text{TT} = -\big[\textsf{A}_{xx}\big]^\text{TT}$ and $\big[\textsf{A}_{xy}\big]^\text{TT} = \big[\textsf{A}_{yx}\big]^\text{TT} = 0$.   Substituting these into the general expression for the transverse-traceless memory wavefunction, as given in Eq.~\eqref{hijtgw-4}, followed by a  decomposition of the wavefunction into $+$ and $\times$ polarization modes,

\begin{eqnarray}\label{eq:hij-mode-decomp}
	\Big[\textsf{h}^\text{mem}_{ij}\Big]^\text{TT} = h_I\,\epsilon^I_{ij} = h^\text{mem}_+\,\epsilon^+_{ij} + h^\text{mem}_\times\,\epsilon^\times_{ij}.
\end{eqnarray}

\noindent allows us to identify:

\begin{eqnarray}\label{eq:wavefunc-ellip}
	h_{+}^\text{mem}(t,\vec x)&=& \cfrac{4G}{r} \int_{-\infty}^{t-r} dt^\prime\cfrac{ 1}{\pi}  \cfrac{G^4 \mu^2 M^3}{a(t^\prime)^5}  \, \times \, \cfrac{2\pi}{15}\, \Big(C_{0}(e) + C_{2}(e) \cos^2 i + C_{4}(e) \cos^4 i\Big), \qquad\qquad h_{\times}^\text{mem}(t,\vec x) = 0.
\end{eqnarray}

\noindent Within the integrand, the time-dependence is encoded in both the semi-major axis $a$ and the eccentricity $e$ of the orbit. This is on account of the fact that energy loss due to primary gravitational waves alters the features of the orbit.
The explicit time-dependence for $a(t)$ and $e(t)$ can be obtained by solving the following system of non-linear differential equations \cite{Peters-2}:
\begin{eqnarray}\label{dadt}
	&& \cfrac{da}{dt} = - \cfrac{64}{5}\,\cfrac{G^{3}\mu M^{2}}{a^{3}\,(1 - e^2)^{7/2}}\,\left( 1 + \cfrac{73}{24}\,e^2 + \cfrac{37}{96}\,e^4 \right),
\end{eqnarray}

\begin{eqnarray}\label{dedt}
&& \cfrac{de}{dt} = - \cfrac{304}{15}\,\cfrac{G^{3}\mu M^{2}}{a^{4}}\,\frac{e}{(1 - e^2)^{5/2}}\left( 1 + \cfrac{121}{304}\,e^2 \right).
\end{eqnarray}
The above can be used to eliminate $t$ and obtain a differential equation involving $a$ and $e$, i.e.,
\begin{eqnarray}\label{eq:a(e)}
\cfrac{da}{de} = \cfrac{12}{19}\,\cfrac{a}{e\,(1-e^2)}\,\cfrac{1 + (73/24)\,e^2 + (37/96)\,e^4}{1 + (121/304)\,e^2},
\end{eqnarray}
whose solution expresses $a$ in terms of $e$,
\begin{eqnarray}
a(e) = \frac{a_{0}}{c_{0}} \,\cfrac{e^{12/19}}{(1-e^2)}\left(1 + \cfrac{121}{304}\,e^2\right)^{870/2299},
\end{eqnarray}
where $c_{0} = \frac{e_{0}^{12/19}}{(1-e_{0}^{2})}(1 + (121/304)\,e_{0}^{2})^{870/2299}$ and $e_{0}$ defines the eccentricity when $a = a_{0}$. The time evolution can be given as a function of eccentricity as
\begin{eqnarray}\label{eq:t-tc(e)}
t - t_{c} = -\cfrac{15}{304}\,\cfrac{a_{0}^{4}}{c_{0}^{4}}\,\cfrac{1}{G^{3}\mu M^{2}}\, \cfrac{19}{48}\,e^{48/19}\,F_{1}\left(\cfrac{24}{19}\,, -\cfrac{1181}{2299}\,, \cfrac{3}{2}\,, \cfrac{43}{19}\,; -\cfrac{121}{304}\,e^2, e^2 \right),
\end{eqnarray}
On the left side of the above equation, $t_c$ corresponds to the instant when the innermost stable circular orbit radius $a_c = 6\,GM$ is reached and on the right side, $F_{1}$ denotes the hypergeometric Appell function with the integral form:
\begin{eqnarray}
F_{1}(a, b_{1}, b_{2}, c; x, y) = \cfrac{\Gamma(c)}{\Gamma(a)\Gamma(c-a)}\int_{0}^{1} t^{a-1}(1-t)^{c-a-1}(1-xt)^{-b_{1}}(1-yt)^{-b_{2}}dt,\,\,\,\,\,\,\,\,  \mathfrak{R}c > \mathfrak{R}a > 0.
\end{eqnarray}
Eq.~\eqref{eq:t-tc(e)} can be expressed entirely in terms of dimensionless quantities by noting that $\overline{T} = \cfrac{a_{0}^{4}}{c_{0}^{4}} \cfrac{1}{(G^{3}\mu M^2)}$ has the dimensions of time. Therefore,
\begin{eqnarray}\label{eq:t-tc(e)-dimless}
	\cfrac{t - t_{c}}{\overline{T}} = \overline{t} - \overline{t}_{c} = -\cfrac{15}{304}\, \cfrac{19}{48}\,e^{48/19}\,F_{1}\left(\cfrac{24}{19}\,, -\cfrac{1181}{2299}\,, \cfrac{3}{2}\,, \cfrac{43}{19}\,; -\cfrac{121}{304}\,e^2, e^2 \right),
\end{eqnarray}
The wavefunction in Eq.~\eqref{eq:wavefunc-ellip} can be rewritten, after a change of integration variable, from $t^\prime$ to $e^\prime$ as,
\begin{eqnarray}\label{eq:h_+(e)}
h_{+}^\text{mem}(e,\vec x) &=& \cfrac{4G}{r}\int_{e}^0 \cfrac{de^\prime}{\dot{e}(e^\prime)}\cfrac{1}{\pi}\cfrac{G^4 \mu^2 M^3}{a(e^\prime)^5}\,\cfrac{2\pi}{15}\left[C_{0}(e^\prime) + C_{2}(e^\prime)\cos^2 i + C_{4}(e^\prime)\cos^4 i \right]\nonumber \\
& = & \cfrac{4G}{r}\,\cfrac{15}{304}\,G \mu M\,\cfrac{c_{0}}{a_{0}}\, \cfrac{2}{15}\, \left[C_{h}^{(0)}(e) + C_{h}^{(2)}(e)\cos^2 i + C_{h}^{(4)}(e)\cos^4 i \right],
\end{eqnarray}
or in terms of dimensionless quantities, we can write
\begin{eqnarray}\label{eq:h_+(e)-dimless}
	\cfrac{h_{+}^\text{mem}(e,\vec x)}{h_0} & = & \overline{h}_{+}^\text{mem}(e,\vec x) = \cfrac{1}{152} \left[C_{h}^{(0)}(e) + C_{h}^{(2)}(e)\cos^2 i + C_{h}^{(4)}(e)\cos^4 i \right],
\end{eqnarray}
where, $h_{0} = \cfrac{4 \,G^2 \mu M}{r}$\,$\cfrac{c_{0}}{a_{0}}$ is the dimension-ful part of the wavefunction. The eccentricity dependent functions $C^{(i)}_h(e)$ are obtained after integrating over the coefficients $C_{i}(e)$, $i = 0,2,4$ given in Eq.~\eqref{eq:C_024-e},
\begin{eqnarray}
C_{h}^{(0)}(e) &=& \cfrac{1}{4}e^{-12/19}\Bigg[ \frac{323}{3}\,\, {}_2F_1 \Bigg(\frac{3169}{2299}, -\frac{6}{19}, \frac{13}{19}; -\frac{121}{304}e^{2} \Bigg) - \frac{29735}{208}\,\, {}_2F_1 \Bigg(\frac{3169}{2299}, \frac{13}{19}, \frac{32}{19}; -\frac{121}{304}e^{2} \Bigg)  \nonumber \\
&&- \frac{29127}{4096}\,\, {}_2F_1 \Bigg(\frac{3169}{2299}, \frac{32}{19}, \frac{51}{19}; -\frac{121}{304}e^{2} \Bigg)  \Bigg]\, , \nonumber \\
C_{h}^{(2)}(e) &=& \frac{1}{4}e^{-12/19}\Bigg[-\frac{304}{3}\, \,{}_2F_1 \Bigg(\frac{3169}{2299}, -\frac{6}{19}, \frac{13}{19}; -\frac{121}{304}e^{2} \Bigg) + \frac{7581}{52}\,\, {}_2F_1 \Bigg(\frac{3169}{2299}, \frac{13}{19}, \frac{32}{19}; -\frac{121}{304}e^{2} \Bigg) \nonumber \\
&& + \frac{1919}{256}\,\, {}_2F_1 \Bigg(\frac{3169}{2299}, \frac{32}{19}, \frac{51}{19}; -\frac{121}{304}e^{2} \Bigg)  \Bigg]\, , \nonumber \\
C_{h}^{(4)}(e) &=& \frac{1}{4}e^{-12/19}\Bigg[-\frac{19}{3}\,\, {}_2F_1 \Bigg(\frac{3169}{2299}, -\frac{6}{19}, \frac{13}{19}; -\frac{121}{304}e^{2} \Bigg) + \frac{2375}{208}\,\, {}_2F_1 \Bigg(\frac{3169}{2299}, \frac{13}{19}, \frac{32}{19}; -\frac{121}{304}e^{2} \Bigg) \nonumber \\
&& + \frac{2071}{4096}\,\, {}_2F_1 \Bigg(\frac{3169}{2299}, \frac{32}{19}, \frac{51}{19}; -\frac{121}{304}e^{2} \Bigg)  \Bigg]\, . \nonumber \\
\end{eqnarray}
The ${}_2F_1$ hypergeometric functions have been identified based on the following integral formulas:
\begin{eqnarray}\label{eq:2F1-function}
	{}_2F_1(a, b, c; z) = \cfrac{\Gamma(c)}{\Gamma(b)\Gamma(c-b)}\int_{0}^{1} x^{b-1}\,(1-x)^{c-b-1}\,(1-zx)^{-a}\,dx, \,\,\,\,\,\,\,\,  \mathfrak{R}c > \mathfrak{R}b > 0.
\end{eqnarray}
The memory waveform has the same structure as reported in Appendix B of \cite{Favata:2011qi}, computed using quadrupole moments.

\begin{figure}[h]
	\includegraphics[scale=0.45]{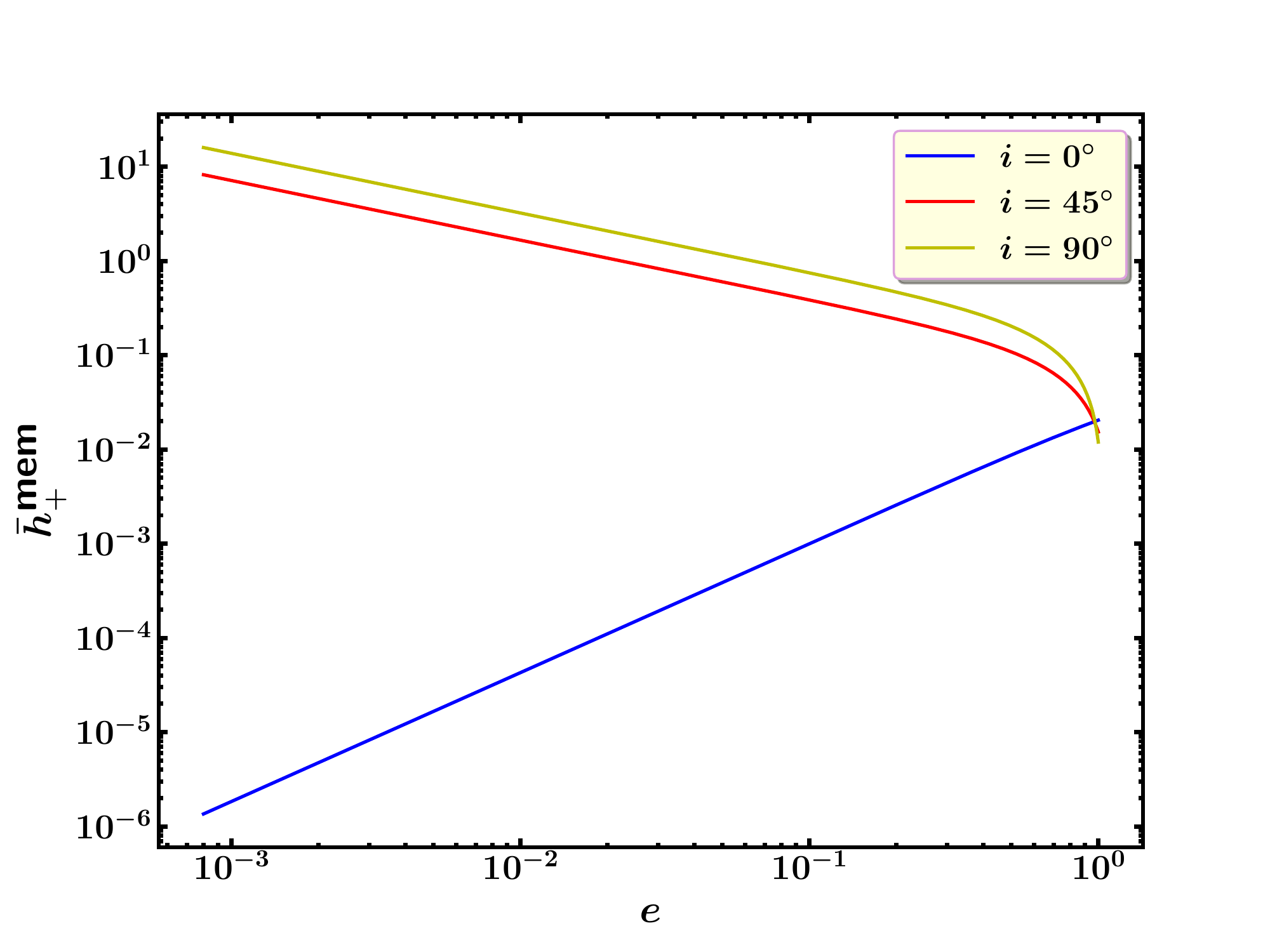}
	\caption{The variation in $\overline{h}_+^\text{mem}$ with respect to the eccentricity of the orbit, plotted for different choices of the angle between the line of sight of observation and the binary axis.}
	\label{fig:h+-mem-vs-ecc}
\end{figure}
\noindent The variation in the memory waveform with varying eccentricity has been elucidated in Fig.~\ref{fig:h+-mem-vs-ecc}. We observe that the gravitational wave has a small but discernible memory for all inclination angles for $e \gg 0$, i.e., long before the moment of reaching the most stable circular orbit. As the orbit plummets to the most stable circular orbit, the memory diminishes for zero inclination, while the memory gradually increases for non-zero inclination as $e$ approaches zero.

\begin{figure}[h]
	\includegraphics[scale=0.45]{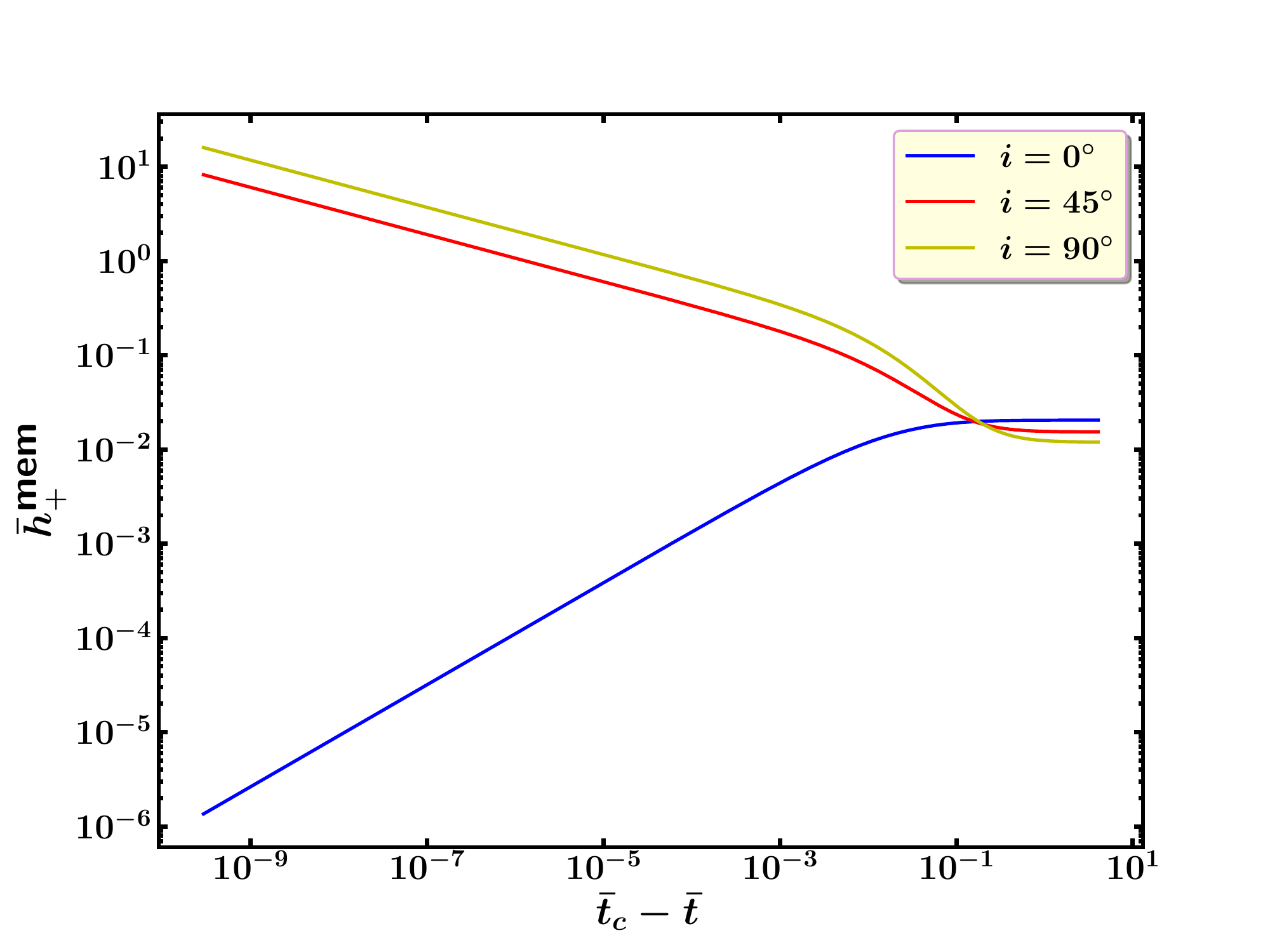}
	\caption{The variation in $\overline{h}_+^\text{mem}$ with respect to the time interval $\overline{t}-\overline{t}_c$ plotted for different choices of the angle between the line of sight of observation and the binary axis.}
	\label{fig:h+-mem-vs-t-tc}
\end{figure}

\noindent The variation with respect to time can be ascertained by utilizing Eq.~\eqref{eq:h_+(e)-dimless} and the implicit relation between time and eccentricity highlighted in Eq.~\eqref{eq:t-tc(e)-dimless}. A plot depicting such a variation is shown in Fig.~\ref{fig:h+-mem-vs-t-tc}. At substantially earlier times than $t_{c}$, memory for different inclination angles saturate at distinct values. As $t$ approaches $t_{c}$, the memory for zero inclination diminishes and eventually reaches an infinitesimal value. While for non-zero inclination angle, memory increases as $t$ approaches $t_{c}$.

\begin{figure}[h]
	\includegraphics[scale=0.45]{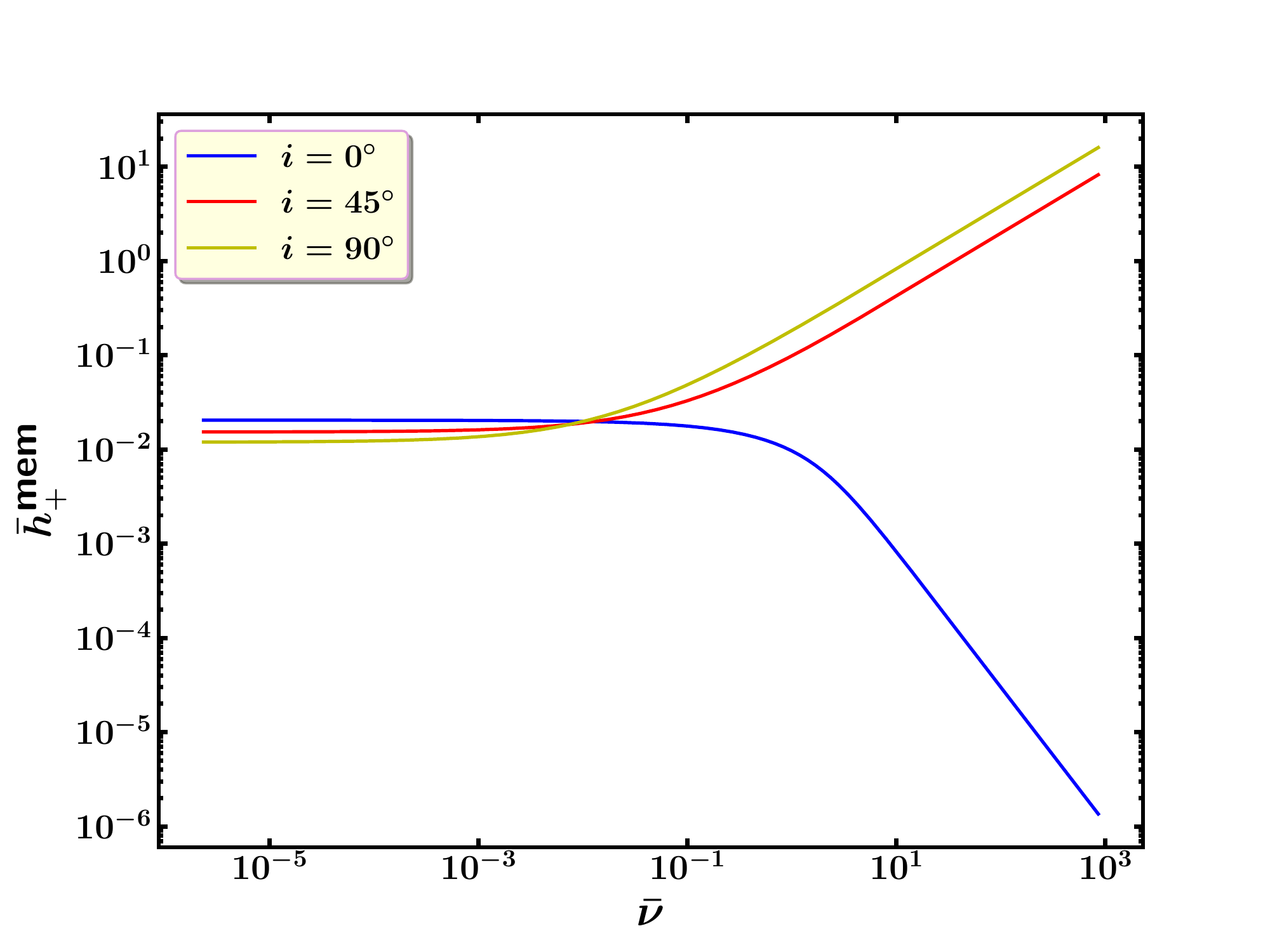}
	\caption{The variation in $\overline{h}_+^\text{mem}$ with respect to the dimensionless frequency parameter $\bar{\nu}$, defined in Eq.~\eqref{eq:nu(e)-dimless},  plotted for different choices of the angle between the line of sight of observation and the binary axis.}
	\label{fig:h+-mem-vs-frequency}
\end{figure}

\noindent We can also describe the behaviour of the memory waveform against the variation in frequency. We first substitute for $a$ in Eq.~\eqref{eq:a(e)} using the relation $\nu = (2\pi)^{-1}\,m^{1/2}\,a^{-3/2}$, which yields the following differential equation: 
\begin{eqnarray}\label{eq:dnu-de}
	\cfrac{d\nu}{de} = -\cfrac{18}{19}\,\cfrac{\nu}{e\,(1-e^2)}\,\cfrac{1 + (73/24)\,e^2 + (37/96)\,e^4}{1 + (121/304)\,e^2}.
\end{eqnarray}
Solving the above we can obtain and expression for the frequency in terms of the eccentricity as,
\begin{eqnarray}\label{eq:nu(e)}
	\nu(e) = \frac{\nu_0}{\widetilde{c}_{0}} \,e^{-18/19}\,\left(1-e^2\right)^{3/2}\left(1 + \cfrac{121}{304}\,e^2\right)^{-1305/2299},
\end{eqnarray}
where $\widetilde{c}_{0} = e_{0}^{-18/19} (1-e_{0}^{2})^{3/2}(1 + (121/304)\,e_{0}^{2})^{-1305/2299}$ with $\nu_{0} = \nu(a_0)$  defining the initial condition.  In terms of dimensionless quantities, the same equation can be rewritten as,
\begin{eqnarray}\label{eq:nu(e)-dimless}
\cfrac{\nu(e)}{\nu_{0}}\,\,\widetilde{c}_{0} = \overline{\nu}(e) =  e^{-18/19}\,\left(1-e^2\right)^{3/2}\left(1 + \cfrac{121}{304}\,e^2\right)^{-1305/2299},
\end{eqnarray}
The variation in the memory signal with respect to the frequency can then be determined by exploiting their mutual dependence on eccentricity, as highlighted in Eqs.~\eqref{eq:h_+(e)-dimless} and \eqref{eq:nu(e)-dimless}. A plot depicting such a relation is shown in Fig.~\ref{fig:h+-mem-vs-frequency}. Eq.~\eqref{eq:nu(e)-dimless} illustrates that as $e$ moves towards $1$, $\Bar{\nu}$ approaches $0$ and  as $e$ goes towards $0$, $\Bar{\nu}$ becomes exceedingly large. It becomes evident from the Fig.~\ref{fig:h+-mem-vs-frequency} that when $\Bar{\nu}$ is very small i.e., the orbit is far away from reaching the most-stable orbit, memory corresponding to different inclinations stay constant at distinct values. The memory for non-zero inclination grows until the most stable circular orbit is reached. On the other hand, the memory falls rapidly for zero inclination as $\Bar{\nu}$ becomes larger than 1.

\section{\bf Conclusions}\label{sec:conclusion}

\noindent In this paper we have computed the linear memory signal for eccentric hyperbolic encounters in both frequency and time domain. We performed a field theoretic calculation of the amplitude of graviton emission from a classical stress-tensor and related this amplitude to the gravitational wave signal as a function of frequency. We took the zero-frequency limit of the waveform to identify the memory signal which we computed in the frequency space and taking the Fourier transform also in the time domain. We performed these calculations for hyperbolic orbits with large eccentricity while retaining terms of all orders in eccentricity. We find that in the eccentric hyperbolic orbits the low-frequency memory component has terms which vary with the frequency $\omega=\nu \Omega$ of the gravitational waves as $\log (\omega e/\Omega)$. This $\log$ dependence is due to the eccentric hyperbolic orbit and is different from the tail terms which occur for any unbound scattering at ${\cal O}(G^2)$. Our results are in agreement with the expressions obtained for the stress-tensors and the memory waveforms based on the soft-graviton theorems \cite{Laddha:2018myi,Sahoo:2018lxl,Laddha:2018vbn,Saha:2019tub}, once appropriate substitutions are made in the latter to describe the specific case of a hyperbolic encounter.

\noindent The non-linear memory which occurs from GW radiated by GW is equally important as it would prove the non-linear nature of graviton-graviton interactions. We have computed the non-linear memory for eccentric elliptical orbits. Highly eccentric orbits are possible if the initial system has three bodies of which one is ejected \cite{Heggie} or by capture from an unbound orbit \cite{Quinlan, OLeary:2008myb}. We compute the frequency spectrum energy radiated by the binaries using the field theory technique which gives the energy spectrum directly in the frequency space. This is used as the source term for the secondary gravitational waves, which results in the memory waveform in the frequency space. We use the instantaneous eccentricity as the parameter for tracking the change of frequency and semi-major axis in time due to radiation reaction. We thus obtain the non-linear memory as a function of the instantaneous eccentricity. We then exploited the change in eccentricity with respect to both time and frequency to obtain the non-linear waveform as a function of both time and frequency. The calculation of the non-linear memory has been done up to all orders in eccentricity. These signal templates in frequency and time domain may be useful for extracting the memory signal from data by the upcoming experiments such as Einstein Telescope and LISA.

In this paper we have used a field-theoretic calculation, and the processes considered are at the tree-level and as expected give the same results as those from the classical quadrupole formula. However, there are some advantages to using the field-theoretic formulation.The energy spectrum is necessarily compared with the sensitivity curves from different experiments to determine the particular experiment for a given type of source. Field theory based methods give us the frequency spectrum of gravitational radiation directly whereas the classical calculation is done in the time domain which then involves an additional Fourier transform.

One specific application of the field theory method is illustrated in this paper where we compare the full orbit calculation with the results from the soft-graviton theorems and found an agreement. This can have applications in the future where calculations of scattering amplitudes can be used for computation of  gravitational waves from astrophysical objects. The field theory formulation is carried out most straightforwardly in the frequency space. The results of gravitational waveforms can be used directly as templates for matching the signals which is carried out in frequency space. This can ease the computational effort in match filtering the signals.

\appendix

\section{Derivation of the general expression for non-linear memory waveforms}\label{app:non-lin-mem-gen-deriv}

\noindent Non-linear memory is due to the secondary gravitational waves which are emitted by the primary gravitational waves from an oscillating source, such as, a coalescing binary \cite{Thorne:1991, Christodoulou:1991, Will:1991, Favata:2008yd, Favata:2010zu, Tolish:2014bka}. 

\noindent The stress-tensor of gravitational waves is related to the energy radiated as
\begin{eqnarray}
	\tau_{ij}^\text{gw}= \cfrac{d E_\text{gw}}{dt \, d\Omega}\, n_i \,n_j , 
\end{eqnarray}
and the non-linear memory waveform is given by \cite{Thorne:1991, Christodoulou:1991,Will:1991},
\begin{eqnarray}
	h_{ij}(t, r \,\hat n)= \cfrac{4G}{r}\,\int_{-\infty}^{t-r} \,dt^\prime \,d\Omega^\prime\, \cfrac{dE_\text{gw}}{dt^\prime \,d\Omega^\prime} \,\left[ \cfrac{n^\prime_i \,n^\prime_j}{1-\vec n^\prime \cdot \vec n}\right]^{\rm TT}.
\end{eqnarray}
Here $ \vec n^\prime$ is the unit vector from the source to the solid angle denoted by $d\Omega^\prime$ and $\vec n$ is the unit vector along the line of sight from the source to the detector.  In what follows, we outline the derivation of these results. 

\noindent The source term of the non-linear gravitational waves is the Issacson stress-tensor of the primary gravitational waves, and it is given as
\begin{eqnarray}
	\tau^\text{gw}_{ij}=\cfrac{1}{32 \pi G} \,\langle \partial_i h^{ab} \partial_j h_{ab} \rangle. 
\end{eqnarray}
Here, the angular brackets encode averaging over time longer than the time period  and volumes larger than the wavelengths of the source gravitational waves  $h_{ab}$. The source gravitational waves travel outward radially with speed of light and are functions of $(t-r)$, i.e $h_{ab}(t, \vec x)=h_{ab}(t-r, \Omega)$. This implies we can relate their spatial and temporal derivatives as,
$\partial_i \,h_{ab}(t-r) =-n_i\, \partial_0\, h_{ab}(t-r)$, where $n_i = x_i/r$. Therefore we can write
\begin{eqnarray}\label{source-1}
	\tau^\text{gw}_{ij}=\cfrac{1}{32 \pi G} \,\langle \partial_i h^{ab} \partial_j h_{ab} \rangle = n_i n_j \tau^\text{gw}_{00}\,.
\end{eqnarray}
We can model the energy density  $\tau^\text{gw}_{00}$ produced from a source and propagating radially outwards on null-rays as 
\begin{eqnarray}
	\tau^\text{gw}_{00}(t, \vec x) = \cfrac{1}{r^2 }\, \cfrac{dE_\text{gw}(t-r, \Omega)}{dt \,d\Omega}.
\end{eqnarray}
where $dE_\text{gw}/dt$ is the luminosity of the source in gravitational waves and $dE_\text{gw}/d\Omega$ denotes the angular distribution of the source luminosity. Therefore we can write $\tau_{ij}^\text{gw}$ in terms of the energy flux as
\begin{eqnarray}
	\tau_{ij}^\text{gw}= n_i n_j \tau_{00}^\text{gw}= n_i n_j  \,\cfrac{1}{r^2 } \, \cfrac{dE_\text{gw}(t-r, \Omega)}{dt \,d\Omega}\,.
\end{eqnarray}
The secondary gravitational waves sourced by the gravitational wave stress-tensor will obey the in-homogenous wave equation:
\begin{eqnarray}\label{wave-secondary}
	\square \,h_{ij}=- 16 \pi G\, \tau^\text{gw}_{ij}.
\end{eqnarray}
whose solution assumes the following form,
\begin{eqnarray}\label{hijtgw-1}
	h^\text{mem}_{ij}(t, \vec x)= 4G \int dt^\prime \,d^3 x^\prime\,\, \tau^\text{gw}_{ij} (t^\prime, \vec x^\prime) \, \cfrac{\delta\left(t^\prime -(t-|\vec x -\vec x^\prime|)\right)}{|\vec x -\vec x^\prime|}.
\end{eqnarray}
We can express the source term, Eq.~\eqref{source-1},  in terms of the null coordinate $u=t^\prime -r^\prime$ as follows,
\begin{eqnarray}\label{tsource-2}
	\tau^\text{gw}_{ij} (t^\prime, \vec x^\prime)&=& \cfrac{n^\prime_i n^\prime_j }{{r^\prime}^2 }\, \cfrac{dE_\text{gw}(t^\prime-r^\prime, \Omega^\prime)}{dt^\prime d\Omega^\prime} \,=\,  \int du \,\,  \cfrac{n^\prime_i n^\prime_j}{{r^\prime}^2 }\, \, \delta(u-(t^\prime-r^\prime)) \,\cfrac{dE_\text{gw}(u, \Omega)}{dt^\prime d\Omega^\prime}.
\end{eqnarray}
Substituting the above in Eq.~\eqref{hijtgw-1} we obtain,
\begin{eqnarray}\label{hijtgw-2}
	h^\text{mem}_{ij}(t, \vec x)= 4G  \,\int  du\,  dt^\prime  \,dr^\prime\, {r^\prime}^2 \,d\Omega^\prime \, \cfrac{n^\prime_i n^\prime_j}{{r^\prime}^2 }\,\cfrac{\delta\left(t^\prime -(t-|\vec x -\vec x^\prime|)\right) }{|\vec x -\vec x^\prime|} \,\, \delta(u-(t^\prime-r^\prime))\,\cfrac{dE_\text{gw}(u, \Omega)}{dt^\prime d\Omega^\prime}. 
\end{eqnarray}
Since the distance to the observer is much larger than the source size, $r \gg r^\prime$, we take the approximations
\begin{eqnarray}
	\cfrac{1}{|\vec x-\vec x^\prime|}\simeq \cfrac{1}{r\, (1-\vec n^\prime \cdot \vec n)}\,,\qquad\qquad \delta\left(t^\prime -(t-|\vec x -\vec x^\prime|)\right)\simeq\delta(t^\prime-(t-r)).
\end{eqnarray}
Now we can perform the integral over $r^\prime$ using the second delta function in Eq.~\eqref{hijtgw-2} and then do the  $t^\prime$ integration using the remaining delta function to obtain
\begin{eqnarray}\label{hijtgw-3}
	h^\text{mem}_{ij}(t, \vec x)= \cfrac{4G}{r} \,  \int_{-\infty}^{t-r} \,du \int_{4\pi}\, d\Omega^\prime\,\,\cfrac{dE_\text{gw}(u, \Omega)}{du\, d\Omega^\prime}\,\, \cfrac{n^\prime_i n^\prime_j }{(1-\vec n^\prime \cdot \vec n)}.
\end{eqnarray}
Expression for the corresponding transverse-traceless wavefunction is obtained after multiplication by the projection operator $\Lambda_{ij,kl}(\vec{n})$, defined in Eq.~\eqref{tensor-1},
\begin{eqnarray}\label{hijtgw-4}
	\Big[h^\text{mem}_{ij}\Big]^{\text{TT}}(t, \vec x) = \cfrac{4G}{r} \,  \int_{-\infty}^{t-r} du \, \int_{4\pi} \, d\Omega^\prime\,\,\cfrac{dE_\text{gw}(u, \Omega)}{du\, d\Omega^\prime}\,\, \cfrac{ \Lambda_{ij,kl}(\vec n)\,  n^\prime_k n^\prime_l }{(1-\vec n^\prime \cdot \vec n)},
\end{eqnarray}

\section{Non-linear memory from circular orbits}\label{app:non-lin-mem-circular}

\noindent As an illustration of the above formalism, we have recomputed the known result for non-linear memory associated with binaries in circular orbits~\cite{Kenneflick-94, Favata:2009ii, Divakarla:2021xrd}.

\noindent Unlike the elliptical case highlighted in section \ref{sec:elliptical}, the circular case only exhibits a single frequency mode. Therefore, the expression for rate of energy radiated in the direction $d\Omega^\prime$ has the following simpler form,
\begin{eqnarray}\label{dEdt-m1}
	\cfrac{dE_\text{gw}}{dt^\prime\, d\Omega^\prime}=\cfrac{ \kappa^2}{4}  \int    \, \Big(T_{ij}(\omega^\prime) \,T^*_{kl}(\omega^\prime)\, \Lambda_{ij,kl}(\vec n^\prime) \Big)\,{\omega^\prime}^3\, 2 \pi\, \delta(\omega^\prime-2 \omega_0) \, \cfrac{d\omega^\prime}{(2 \pi)^3 \, 2 \omega^\prime}.
\end{eqnarray}
where $\omega_0=\sqrt{G(m_1+m_2)/a^3}$ is the angular frequency of the Kepler orbit. The stress-energy tensor $T_{ij}$, where $i,j = x,y,z$, can be written in matrix form as:
\begin{eqnarray}
	\textsf{T}\, (\omega_0) = \cfrac{\mu\, a^2 \omega_0^2}{2}\,\begin{pmatrix}
		1 && \iota && 0\\
		\iota && -1 && 0\\
		0 && 0 && 0
	\end{pmatrix}.
\end{eqnarray}

\noindent We consider the orientation defined in Fig.~\ref{fig:Non-Linear-Memory}, where the axis of rotation of the binary system is counter-clockwise rotated, by an angle $i$, with respect to the $z$ axis. This rotation impacts the stress-tensor matrix \textsf{T} as follows: 
\begin{eqnarray}\label{eq:rotation}
	\textsf{T} \rightarrow \textsf{T}^\prime = \mathcal{R}\,\textsf{T}\,\mathcal{R}^\text{T}, \hspace{0.2cm} \text{with} \hspace{0.2cm} \mathcal{R} =  \begin{pmatrix}
		1 && 0 && 0\\
		0 && \cos i && -\sin i\\
		0 && \sin i && \cos i
	\end{pmatrix}, \hspace{0.1cm} \Rightarrow \hspace{0.1cm} \textsf{T}^\prime = \cfrac{\mu\, a^2 \omega_0^2}{2} \begin{pmatrix}
		1 && \iota \cos i && \iota \sin i\\
		\iota \cos i && -\cos^2 i && -\cos i\sin i\\
		\iota \sin i && -\cos i\sin i && -\sin^2 i
	\end{pmatrix} .
\end{eqnarray}

\noindent The product $T_{ij}\,T^*_{kl}\,\Lambda_{ij,kl}(\vec n^\prime)$ which can be rewritten entirely as a product of $T_{ij}$'s and $\hat{n}^\prime$, using Eq.~\eqref{tensor-1}, as follows:
\begin{eqnarray}\label{eq:tensor-product-index}
	T_{ij}\,T^*_{kl}\,\Lambda_{ij,kl}(\hat n^\prime) &=& T_{ij}\,T^*_{ji} - 2\, T_{ij}\,T^*_{jl}\,n_i^\prime\, n_l^\prime  + \cfrac{1}{2}\,T_{ij}\, T^*_{kl}\, n_i^\prime\,  n_j^\prime\,  n_k^\prime\,  n_l^\prime   + \cfrac{1}{2}\,\left( T_{ii}\,T^*_{kl}\, n_k^\prime\, n_l^\prime + T_{ij}\,T^*_{kk}\, n_i^\prime\, n_j^\prime - T_{ii}\,T^*_{kk} \right).
\end{eqnarray}

\noindent For the circular case both the stress-tensor matrices $\mathsf{T}$ and $\mathsf{T}^\prime$ are already traceless. Therefore, the last three terms, within the parentheses, on the right hand side of Eq.~\eqref{eq:tensor-product-index} vanish. The remaining terms can be evaluated as products of matrices as shown below:
\begin{eqnarray}\label{eq:tensor-product-matrix}
	T_{ij}\,T^*_{kl}\,\Lambda_{ij,kl}(\hat n^\prime) \,&\equiv&\, \text{Tr}\,[\mathsf{T}^\prime\,\mathsf{T}^{\prime\dagger}] - 2\,\left(\hat{n}^{\prime\text{T}}\,\mathsf{T}^\prime\right)\cdot\,\left(\mathsf{T}^{\prime\dagger}\hat{n}^\prime\right) + \cfrac{1}{2}\,\left(\hat{n}^{\prime\text{T}}\,\mathsf{T}^\prime\hat{n}^\prime\right)\,\left(\hat{n}^{\prime\text{T}}\,\mathsf{T}^{\prime\dagger}\hat{n}^\prime\right)\,.
\end{eqnarray}
Substituting for $\mathsf{T}^\prime$, $\hat{n}^\prime$ in Eq.~\eqref{eq:tensor-product-matrix} and evaluating the matrix products and traces yields:
\begin{eqnarray}\label{eq:tensor-product-angles}
	T_{ij}\,T^*_{kl}\,\Lambda_{ij,kl}(\hat n^\prime) \,&\equiv&\, \cfrac{\mu^2 a^4 \omega_0^4}{4}\, \cfrac{1}{2}\,\Big(1+ 6 \cos^2\theta + \cos^4\theta \Big)\,.
\end{eqnarray}
Here, $\theta$ is the angle between the axis of rotation of the binary  $\vec{L} = \left(0,-\sin i, \cos i \right)$ and the direction where the primary graviton emits the second graviton: $\hat n^\prime= \left(\sin \theta^\prime \cos\phi^\prime, \sin\theta^\prime \sin \phi^\prime, \cos\theta^\prime \right)$, i.e.,
\begin{eqnarray}
	\cos \theta = -\sin i \sin \theta^\prime \sin \phi^\prime + \cos i \cos \theta^\prime.
\end{eqnarray} 
Eq.~\eqref{dEdt-m1} can now be expressed, using the result of Eq.~\eqref{eq:tensor-product-angles} as
\begin{eqnarray}
	\cfrac{dE_\text{gw}}{dt^\prime\, d\Omega^\prime}=\cfrac{G}{2\pi}\,\omega_0^6\, \mu^2 a^4\, \Big(1+ 6 \cos^2\theta + \cos^4\theta \Big)\,.
\end{eqnarray}
The Kepler orbital frequency can be written in terms of the semi-major axis $a$, i.e. $\omega_0=(G M/ a^3)^{1/2}$ ($M=m_1+m_2$). This allows us to rewrite the expression for the rate of energy radiated as,
\begin{eqnarray}
	\cfrac{dE_\text{gw}}{dt^\prime d\Omega^\prime}=\cfrac{G^4}{2\pi}\cfrac{ \mu^2 M^3}{ a^5} \Big(1+ 6 \cos^2\theta + \cos^4\theta \Big)\,.
\end{eqnarray}
Substituting the above in Eq.~\eqref{hijtgw-3}, we obtain the following expression for the memory signal:

\begin{eqnarray}\label{hijtgw-5}
	h_{ij}^\text{mem}(t, \vec x)&=& \cfrac{4G}{r}   \int_{-\infty}^{t-r} dt^\prime \int_{4\pi} d\Omega^\prime\,\cfrac{dE_\text{gw}(t^\prime, \Omega)}{dt^\prime d\Omega^\prime} \cfrac{n^\prime_i n^\prime_j }{(1-\hat n^\prime \cdot \hat n)}\nn\\
	&=&  \cfrac{4G}{r}   \int_{-\infty}^{t}  dt^\prime\, \cfrac{G^4}{2\pi}\,\cfrac{ \mu^2 M^3}{ a^5}  \underbrace{\int_{4\pi} d\Omega^\prime\,\Big(1+ 6 \cos^2\theta + \cos^4\theta \Big) \cfrac{n^\prime_i n^\prime_j }{(1-\hat n^\prime \cdot \hat n)}}_{\textsf{A}_{ij}}.
\end{eqnarray}

\noindent where we have replaced $t-r$ by $t$ in the limit of the integrand as the waves are detected at a fixed $r$ and the dependence in $u$ then becomes a dependence on time of observation $t$. \\
\noindent To obtain the transverse-traceless part of $h^\text{mem}_{ij}$ we apply the TT projection operator, through multiplication with $\Lambda_{ij,kl}(\vec{n})$, see Eq.~\eqref{hijtgw-4},

\begin{eqnarray}\label{hijtgw-6}
	\Big[h_{ij}^\text{mem}\Big]^\text{TT}(t,\vec x)&=& \cfrac{4 G^5}{2 \pi r}\, \mu^2 M^3  \int_{-\infty}^{t} dt^\prime\cfrac{ 1}{ a^5}  \, \underbrace{\int_{4\pi} d\Omega^\prime\,\Big(1+ 6 \cos^2\theta + \cos^4\theta \Big) \cfrac{ \Lambda_{ij,kl}(\hat n)\, n^\prime_k n^\prime_l }{(1-\hat n^\prime \cdot \hat n)}}_{\big[\textsf{A}_{ij}\big]^\text{TT}}.
\end{eqnarray}
%
Substituting for the components of $\hat{n}$, $\hat{n}^\prime$ and $\delta_{ij}$, the angular integrals can be evaluated, for $i,j = x,y$ as follows:

\begin{eqnarray}\label{Eq:CircA11}
	\Big[\textsf{A}_{xx}\Big]^\text{TT} &=& \int_{0}^{\pi} \sin\theta^\prime \,d\theta^\prime\, \int_{0}^{2\pi} d\phi^\prime \,\left(1+ 6 \cos^2\theta + \cos^4\theta \right)\,\times\,\cfrac{1}{2}\left( 1 + \cos \theta^\prime\right)\,\cos 2\phi^\prime = -\cfrac{2\pi}{15}\,\left(17 + \cos^2 i\right) \sin^2 i, \nonumber\\
	\Big[\textsf{A}_{xy}\Big]^\text{TT} &=& \int_{0}^{\pi} \sin\theta^\prime \,d\theta^\prime\, \int_{0}^{2\pi} d\phi^\prime \,\left(1+ 6 \cos^2\theta + \cos^4\theta \right)\,\times\,\cfrac{1}{2}\left( 1 + \cos \theta^\prime\right)\,\sin 2\phi^\prime = 0.
\end{eqnarray}

\noindent The other components are: $	\Big[\textsf{A}_{yy}\Big]^\text{TT} = -\Big[\textsf{A}_{xx}\Big]^\text{TT}$ and $\Big[\textsf{A}_{xy}\Big]^\text{TT} = \Big[\textsf{A}_{yx}\Big]^\text{TT} = 0$. The transverse-traceless gravitational wavefunction can be decomposed into $+$ and $\times$ modes, see Eq.~\eqref{eq:hij-mode-decomp}. We can then identify,

\begin{eqnarray}\label{eq:h-mem-plus-1}
	h_{+}^\text{mem}(t,\vec x)&=& \cfrac{4 G^5}{2 \pi r}\,\, \mu^2 M^3  \int_{-\infty}^{t-r} dt^\prime\cfrac{ 1}{ a^5}  \, \times \, \left[\cfrac{2\pi}{15}\,\left(17 + \cos^2 i\right) \sin^2 i \right], \qquad\qquad h_{\times}^\text{mem}(t,\vec x) = 0.
\end{eqnarray}

\noindent time dependence of the integrand in Eq.~\eqref{hijtgw-5} is due to the change in the radius of the orbit which occurs
due to the energy loss of the orbit by the primary gravitational waves.  This change in the radius is given as  $da/dt= (dE/dt)\,(da/dE)$, which using $E=-(1/2)\,G \mu M/a $ is

\begin{eqnarray}\label{dadtbinary1}
	\cfrac{da}{dt} =-\cfrac{64\, G^3}{5}\, \cfrac{\mu M^2}{a^3}\,.
\end{eqnarray}

\noindent Solving this equation gives us the time dependence of the separation distance $a(t)$

\begin{eqnarray}\label{at-1}
	a(t)=a_{c}\left[1 + \cfrac{256}{5} \, \cfrac{G^3 \mu M^2}{a_{c}^{4}}  (t_c-t) \right]^{1/4}.
\end{eqnarray}

\noindent where $t_c$ is the time when the innermost stable circular orbit radius $a_c = 6\,GM$ is reached\footnote{The physics beyond this time cannot be modelled within the framework of Keplerian dynamics employed throughout our calculations.}. The frequency $f=\omega_0/\pi = (1/\pi) \sqrt{(GM)/a^3}$ increases till $t \rightarrow t_c$

\begin{eqnarray}
	f(t^\prime) &\simeq &  \left(\cfrac{5}{256}\right)^{3/8}\cfrac{1}{ \pi}\, (G {\mathcal M_c})^{-5/8} \left[\cfrac{5}{256}\, \cfrac{a_{c}^{4}}{G^{3}\mu M^2} + (t_{c} - t^{\prime}) \right]^{-3/8}
\end{eqnarray}

\noindent where ${\cal M}_c= \mu^{3/5}M^{2/5}$ is the chirp mass of the binary pair. Using Eq.~\eqref{at-1} to substitute for $a(t)$ in Eq.~\eqref{eq:h-mem-plus-1}, we obtain the non-linear memory wavefunction of binaries, whose rotation axis makes an angle $i$ with respect to the earth-source direction, as 

\begin{eqnarray}\label{{eq:h-mem-plus-1}}
	h^\text{mem}_+(t) & = & \cfrac{1}{192}\cfrac{G}{r}\left(5 G\mu^3 M^{2}\right)^{1/4}\, \frac{1}{\left[5 \left(\frac{3}{2}\right)^4 \frac{GM^2}{\mu} + (t_c - t)\right]^{1/4}} \, {\sin^2 i \left(17+\cos^2 i\right)}\nonumber \\
	& = & h_{\text{circ}}\, \cfrac{1}{\left[1 + (t^{\prime}_{c} - t^{\prime}) \right]^{1/4}}\, {\sin^2 i \left(17+\cos^2 i\right)},
\end{eqnarray}

\noindent where $h_{\text{circ}} = \frac{2}{3}\, \frac{1}{192}\, \frac{G\mu}{r}$ is the dimension-ful part of the waveform and $(t^{\prime}_{c} - t^{\prime}) = (t_{c} - t)/T^{\prime}$ with $T^{\prime}$ being $\left(\frac{3}{2}\right)^{4}\frac{5GM^2}{\mu}$ has the units of time. The memory waveform can also be expressed in frequency domain and has the form

\begin{eqnarray}
h^\text{mem}_+(\omega_{0}) = (6 a_{c}^{2})^{1/3}\,h_{\text{circ}}\,\omega_{0}^{2/3}.
\end{eqnarray}

\begin{figure}[!htb]
	\includegraphics[scale=0.65]{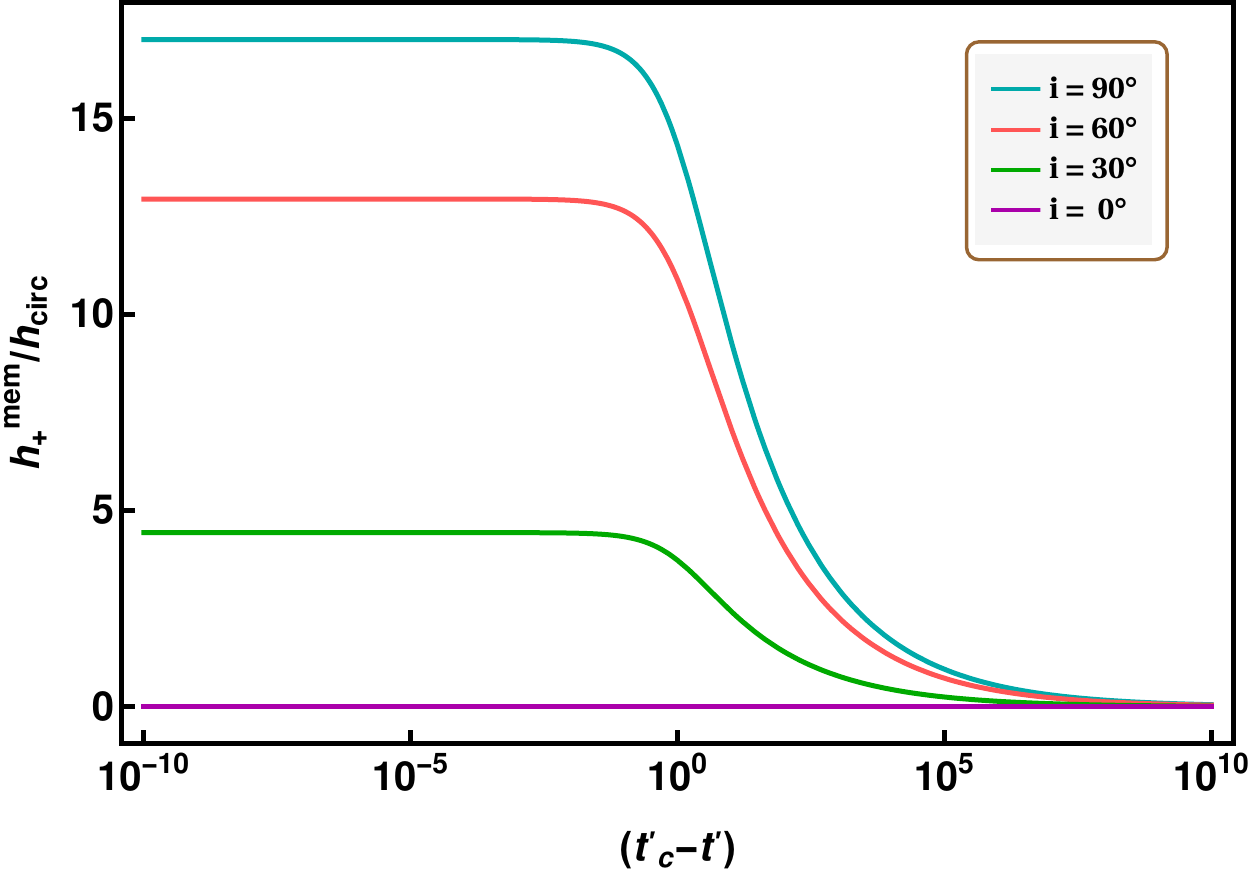}
	\caption{The variation in $h_+^\text{mem}$ against time plotted for different choices of the angle $i$ between the line of sight of observation and the binary axis.}
	\label{fig:h-mem-vs-t-circular}
\end{figure}

\noindent We have demonstrated the dependence of the waveform on the angle $i$ in Fig.~\ref{fig:h-mem-vs-t-circular}. An earlier work~\cite{Kenneflick-94} considered the orientation of the binary rotation axis to lie in the $x-z$ plane, while in this work, the binary rotation axis lies in the $y-z$ plane (see Fig.~\ref{fig:Non-Linear-Memory}). The memory waveforms exhibit exactly the same behaviour as shown in Fig.~1 of Ref.~\cite{Kenneflick-94}. To understand the relation between the individual plots in Fig.~\ref{fig:h-mem-vs-t-circular} with the physical picture of an inspiralling binary, they must be read from right to left. The rightmost points on the curve represent vanishing memory signal irrespective of the angle $i$, for very early times. As $t^\prime \rightarrow t_c^\prime$ or as $t_c^\prime - t^\prime$ becomes small, the memory signal grows and eventually saturates at a maximum value around $t_c^\prime - t^\prime \sim T^\prime = \left(\frac{3}{2}\right)^{4}\frac{5GM^2}{\mu}$ which is the time of caoalescence of the binary to form a single black-hole. The memory effect will be strongest in edge on binaries $i=\pi/2$ and will be zero in the face on binaries $i=0$. 

\section{Formulae for summing over products of Bessel functions}\label{app:sum-formulas}

For evaluating the sum over $n$ for elliptical binaries, we have used following formulae \cite{Peters-1}:
\begin{eqnarray}\label{eq:Peter-Matthews-sum}
&& \sum_{n=0}^{\infty} \, n^{2}J_{n}^{2}(ne) = \cfrac{e^{2}}{4(1-e^{2})^{7/2}}\left(1 + \cfrac{e^{2}}{4} \right), \nonumber \\
&& \sum_{n=0}^{\infty} \, n^{2}[J_{n}^{\prime}(ne)]^{2} = \cfrac{1}{4(1-e^{2})^{5/2}}\left(1 + \cfrac{3e^{2}}{4} \right), \nonumber \\
&& \sum_{n=0}^{\infty} \, n^{3}J_{n}(ne) J_{n}^{\prime}(ne) = \cfrac{e}{4(1-e^{2})^{9/2}}\left(1 + 3e^{2} + \cfrac{3}{8}e^{4} \right), \nonumber \\
&& \sum_{n=0}^{\infty} \, n^{4}J_{n}^{2}(ne) = \cfrac{e^{2}}{4(1-e^{2})^{13/2}}\left(1 + \cfrac{37}{4}e^{2} + \cfrac{59}{8}e^{4} + \cfrac{27}{64}e^{6} \right), \nonumber \\
&& \sum_{n=0}^{\infty} \, n^{4}[J_{n}^{\prime}(ne)]^{2} = \cfrac{1}{4(1-e^{2})^{11/2}}\left(1 + \cfrac{39}{4}e^{2} + \cfrac{79}{8}e^{4} + \cfrac{45}{64}e^{6} \right).
\end{eqnarray}

\begin{acknowledgements}

\noindent The authors would like to thank Ashoke Sen and Biswajit Sahoo for helpful discussions. They would also like to thank Diptarka Das for hosting S.M. at the Indian Institute of Technology Kanpur, which was supported through the Max Planck Partner Group grant MAXPLA/PHY/2018577. S.P. is supported by the MHRD, Government of India, under the Prime Minister's Research Fellows (PMRF) Scheme, 2020. A.H. would like to thank the MHRD, Government of India for research fellowship.  
\end{acknowledgements}

	\newpage
	\appendix
%
%
	
\end{document}